\documentclass[apjl,twocolumn]{emulateapj_mod}
\usepackage{epsfig,apjfonts,mathptmx}

\def\gtsima{$\; \buildrel > \over \sim \;$}
\def\ltsima{$\; \buildrel < \over \sim \;$}
\def\prosima{$\; \buildrel \propto \over \sim \;$}
\def\gsim{\lower.5ex\hbox{\gtsima}}
\def\lsim{\lower.5ex\hbox{\ltsima}}
\def\simgt{\lower.5ex\hbox{\gtsima}}
\def\simlt{\lower.5ex\hbox{\ltsima}}
\def\simpr{\lower.5ex\hbox{\prosima}}

\def\h1{$h^{-1}$}
\def\mguv{$Mg_{UV}$\ }
\def\eeq{\end{equation}}
\def\beq{\begin{equation}}
\def\spb{\smallskip\par\noindent$\bullet\;$}

\shorttitle{Passive $z>1.4$ galaxies in the UDF}
\shortauthors{E. Daddi et al.}

\begin{document}

\title{Passively Evolving Early-type Galaxies at
$1.4\simlt\lowercase{z}\simlt 2.5$ in the Hubble Ultra Deep Field}
\author{E. Daddi\altaffilmark{1,2},
	A. Renzini\altaffilmark{2},
	N. Pirzkal\altaffilmark{3},
	A. Cimatti\altaffilmark{4},
	S. Malhotra\altaffilmark{3},
	M. Stiavelli\altaffilmark{3},
	C. Xu\altaffilmark{3},
	A. Pasquali\altaffilmark{5},
	J.~E. Rhoads\altaffilmark{3},
	M. Brusa\altaffilmark{6},
	S. di Serego Alighieri\altaffilmark{4},
	H.~C. Ferguson\altaffilmark{3},
	A.~M. Koekemoer\altaffilmark{3},
	L.~A. Moustakas\altaffilmark{3},
	N. Panagia\altaffilmark{3},
	R.~A. Windhorst\altaffilmark{7}
}

\altaffiltext{$\star$}{Based on observations with the NASA/ESA {\em
Hubble Space Telescope}, obtained at the Space Telescope Science
Institute, which is operated by AURA Inc, under NASA contract NAS
5-26555; also based on data collected at the European
Southern Observatory, Chile.}  
\altaffiltext{1}{{\em Spitzer Fellow}, National Optical Astronomy Observatory, 
P.O. Box 26732, Tucson, AZ 85726, USA   -- edaddi@noao.edu
}
\altaffiltext{2}{European Southern Observatory,
Karl-Schwarzschild-Str. 2, D-85748 Garching, Germany
} 
\altaffiltext{3}{Space Telescope Science Institute, 3700 SanMartin
Drive, Baltimore, MD21218, USA}
\altaffiltext{4}{INAF--Osservatorio Astrofisico di
Arcetri, L.go E. Fermi 5, Firenze, Italy} 
\altaffiltext{5}{Institute of Astronomy, ETH H\"{o}nggerberg, 8093
Zurich, Switzerland}
\altaffiltext{6}{Max Planck Institut fuer Extraterrestrische Physik,    D-85478 Garching, Germany}
\altaffiltext{7}{Dept. of Physics \& Astronomy, Arizona State
University, P.O. Box 871504, Tempe, AZ
85287-1504, USA}

\begin{abstract}
We report on a complete sample of 7 luminous
early-type galaxies in the Hubble Ultra
Deep Field (UDF) with spectroscopic
redshifts between 1.39 and 2.47 and to $K_{AB}<23$.  
Using the $BzK$
selection criterion we have pre-selected a set of
objects over the UDF which fulfill the photometric conditions for being
passively evolving galaxies at $z>1.4$. Low-resolution spectra of
these objects have been extracted from the HST+ACS grism data taken over
the UDF by the GRAPES project. Redshift for the 7
galaxies have been identified based on the UV feature at rest frame
$2640<\lambda<2850$\ \AA. This feature is 
mainly due to a combination of FeII, MgI and
MgII absorptions which are characteristic of stellar populations
dominated by stars older than $\sim 0.5$ Gyr. The redshift identification
and the passively evolving nature of these galaxies is further
supported by the photometric redshifts and by the overall spectral energy
distribution (SED), with the ultradeep HST+ACS/NICMOS imaging revealing
compact morphologies typical of elliptical/early-type galaxies. From
the SED we derive stellar masses of $\simgt 10^{11}M_\odot$ and ages of
$\sim 1$ Gyr.  Their space
density at $<\! z\! >=1.7$ appears to be roughly a factor of 2--3 smaller
than that of their local counterparts, further supporting the notion
that such massive and old
galaxies are already ubiquitous at early cosmic times.
Much smaller
effective radii are derived for some of the objects compared to local
massive ellipticals, which may be due to morphological K corrections, 
evolution, or the presence of a central point-like source. Nuclear activity
is indeed present in a subset of the galaxies, as revealed by them being
hard X-ray sources, hinting to AGN activity having played a role in 
discontinuing star formation.
\end{abstract}
\keywords{
galaxies: evolution --- 
galaxies: formation --- 
galaxies: early-types --- 
galaxies: high-redshift ---
cosmology: observations
}

\section{Introduction}

In the local universe,
as accurately measured by the SDSS (e.g., Baldry et al. 2004), 
passive early type galaxies with stellar masses larger than
$10^{11}M_\odot$ dominate the counts of most massive 
galaxies, being a factor of 3 more numerous than late type galaxies
above this mass threshold.
About 1/3 of all the stars in galaxies in the local universe are hosted
by such objects (Baldry et al. 2004). 
The process by which these galaxies formed is still unclear.
At least some of them may have formed at relatively high redshifts,
in a process 
that, for its rapidity, is reminiscent of the monolithic collapse scenario 
(Eggen, Lynden-Bell \& Sandage 1962).
On the other hand, others may have been assembled at relatively
recent epochs, through merging of smaller subunits (Toomre 1977).
The formation of massive spheroids is a central problem of current
theories of galaxy formation (e.g., Loeb \& Peebles 2003; Gao et al.
2004).

It is now well established that up to $z\sim1$ a significant population
of red passively evolving early-type galaxies can be found 
in the field among extremely red objects (EROs;
Cimatti et al. 2002a; Yan et al. 2004; see McCarthy et al. 2004 for a
review), together with dust reddened
systems. The EROs are highly
clustered (Daddi et al. 2000a; McCarthy et al. 2001; Roche et al.
2003; Miyazaki et al. 2003; Brown et al. 2005; Georgakakis et al. 2005) as expected for
the progenitors of local massive ellipticals (Daddi et al. 2001;
Moustakas \& Somerville 2002). The space density of 
$z\sim1$ early-type galaxies is at least within a factor of 2 of the
local value (Daddi et al. 2000b; Pozzetti et al. 2003; Bell et al. 2004;
Caputi et al. 2004)
implying a modest evolution from $z=1$, especially for
the most massive ones.
Their stars appear fairly old ($\simgt3$\ Gyr), suggesting even higher 
formation redshifts $z\simgt2$ (Cimatti et al. 2002a; Treu et al. 2005). 

On the other hand, physically motivated galaxy formation models based
on hierarchical clustering within the $\Lambda$CDM framework
(Cole et al. 2001; Kauffmann et al. 1999;
Somerville et al. 2001) have been so far unable to account for the
large space density of $z>1$ red galaxies (Daddi et al. 2000b; Smith
et al. 2001; Firth et al. 2002; Somerville et al. 2004; Glazebrook et
al. 2004). These models 
result in widespread merging and associated star-formation activity of
massive galaxies at relatively low-redshifts, and it is currently not
fully clear what physical mechanism needs to be involved in order
to terminate star-formation and produce the red colors of passive 
galaxies. Feedback processes, e.g., from the onset of AGN activity,
seem to be a promising tool to achieve that
(Granato et al. 2001; 2004; Springel et 
al. 2004).
This strengthens the necessity to simultaneously trace the formation
of galaxies and AGN activity 
to understand the link between assembly of stellar mass in galaxies 
and the growth of supermassive black holes
(Magorrian et al. 1998; Ferrarese \& Merritt 2000).

As little evolution in massive ($>10^{11}M_\odot$) field
early-type galaxies is detected up to
$z=1$, it is necessary to push the investigation to the highest possible
redshifts in order to further constrain the formation of early type galaxies. 
Crucial questions to ask are: up to which redshift do passively
evolving early-type galaxies exist ?; which is their space density
as a function of $z$ ?; what is their clustering ? and what is the
environment they live in (i.e. cluster vs field) ?
Studying the highest redshift massive and passive objects, i.e.
those presumably
closest to the formation or assembly epoch, could in principle also reveal
useful information to understand the physical processes
by which these galaxies were formed.

For almost a decade the unique example of spectroscopically confirmed 
high-z passive galaxy beyond $z\sim1.5$ has been 53W091 at $z=1.55$
(Dunlop et al. 1996; Spinrad et al. 1997). This objects was preselected
for being a radio-galaxy, hence it was virtually selected over the whole 
sky. Its very red colors suggested fairly old stellar populations, 
although with some controversies about detailed age (see e.g., Nolan et al. 2001
and references therein).
The North and South Hubble Deep Fields with their extremely deep imaging
data
prompted searches for $z>1.5$ passive galaxy candidates (Treu et al.
1999; Stiavelli et al. 1999; Broadhurst \& Bowuens 2000; Benitez et al.
1999), however none of such objects was eventually spectroscopically 
confirmed. That is due to the faintness of the candidates
in the optical domains
and because optical spectroscopy from the ground 
is hampered by the presence of strong OH sky emission lines at the wavelengths
where the main spectral features are redshifted.
Now, over fields of a few tens to a few hundreds arcmin$^2$,
three groups have recently presented the discovery of several
high-$z$ passive galaxies.
Cimatti et al. (2004) reports 4 passive galaxies spectroscopically
confirmed  at
$1.6<z<1.9$ in the K20 survey, from a region within the GOODS-South field. The ACS
imaging revealed their compact early-type galaxy like morphologies.
McCarthy et al. (2004) have redshifts for 20 passive objects in 
$1.3<z<2.15$ (of which 8 at $z>1.5$) from the 
Gemini Deep Deep Survey. Saracco et al. 2005
confirm 7 very massive old galaxies at $1.3<z<1.7$ from low-resolution near-IR 
spectroscopy. The latter two studies both lack HST 
morphology information.

While it is now established that a significant population of passive galaxies
exist up to $z\sim2$ at least, the availability of the
Hubble Ultra Deep Field (UDF) dataset,
with ultimately deep and high spatial resolution optical and near-IR imaging
offers a unique possibility for studying these high-z passive 
galaxies in some detail. 
As a part of our Grism ACS Program for Extragalactic Science (GRAPES), 
we have collected about 
50 HST orbits of ACS+G800L grism spectroscopy on the UDF
(Pirzkal et al. 2004, P04 hereafter).  As discussed in more detail 
further in the paper, the HST+ACS spectroscopy allows us to obtain much higher
S/N ratios on the continuum than reachable from the ground. 
In this paper, we have taken advantage of the GRAPES spectra to present
spectroscopic confirmation for a sample of 7 early-type galaxies at
$1.4<z<2.5$ in the UDF, and used the available multiwavelength datasets 
to study their properties.

The paper is organized as follows: in Sect.~\ref{sect:MgUV}
we comment on the usefulness of low
resolution spectra of passive galaxies and a new spectral
index is defined for their characterization. Sect.~\ref{sect:BzK}
discusses the color selection of $z>1.4$ passive galaxy candidates.
In Sect.~\ref{sect:reds}
redshift identifications are presented and discussed in detail. 
Morphological parameters of the 7 objects are derived in
Sect.~\ref{sec:galfit}. Stellar population properties, including stellar
masses and ages, are estimated in Sect.~\ref{sec:fits}. The issue of
space density evolution of passive early-type galaxies 
to $z=2.5$ is discussed in
Sect.~\ref{sec:evol}, while in Sect.~\ref{sec:mor_evol} the evolution of
morphology is discussed. Sect.~\ref{sec:Xrays}
presents the X-ray detection of two of the galaxies in our sample.
Summary and conclusions
are given in Sect.~\ref{sec:summary}.

We assume a Salpeter IMF from 0.1 and 100 $M_\odot$,
and a WMAP flat cosmology with
$\Omega_\Lambda, \Omega_M = 0.73, 0.27$, and
$h = H_0$[km s$^{-1}$ Mpc$^{-1}$]$/100=0.71$ (Spergel et al. 2003).

\begin{figure*}[ht]
\centering 
\includegraphics[width=16cm]{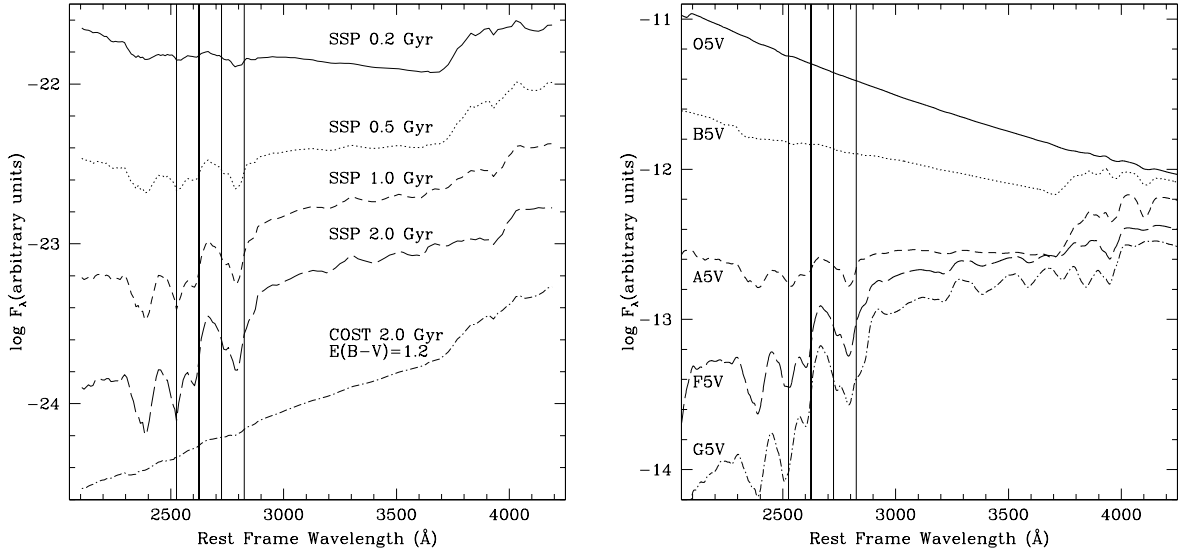}
\caption{
The mid-UV spectral energy distribution 
of galaxies from the
Bruzual \& Charlot (2003) library (left) 
and of stars from the Kurucz database (right), 
smoothed to $\sim50$ \AA\ rest frame
resolution, similar to the HST spectra of the objects
analysed in this paper.
Old simple stellar 
population (SSP) galaxies or low mass stars
have very red spectra and a characteristic signature with a peak and a dip 
in the region 
2640--2850\AA, mainly due to a combination of FeII, MgI and
MgII absorptions lines.
This $Mg_{UV}$ feature
appears after a few 100 Myr of passive evolution (visible in A-type
stars or later) and gets stronger as 
age increase (notice the 2 order of magnitude UV continuum dimming
for SSP galaxies from 0.2 to 2.0 Gyr).
Star-forming galaxies with
large reddening (dot-dashed line in the left panel)
can reach similarly red continua in case of very high
reddening, but are distinguished from the former because of the lack
of any strong spectral feature in the UV, like OB-type stars.
The vertical lines (both panels)
show the 3 windows used  for the measurement of the
$Mg_{UV}$ index (Eq.~\ref{eq:MgUV}).
}
\label{fig:MgUV1}
\end{figure*}

\begin{figure}[ht]
\centering 
\includegraphics[width=8.8cm]{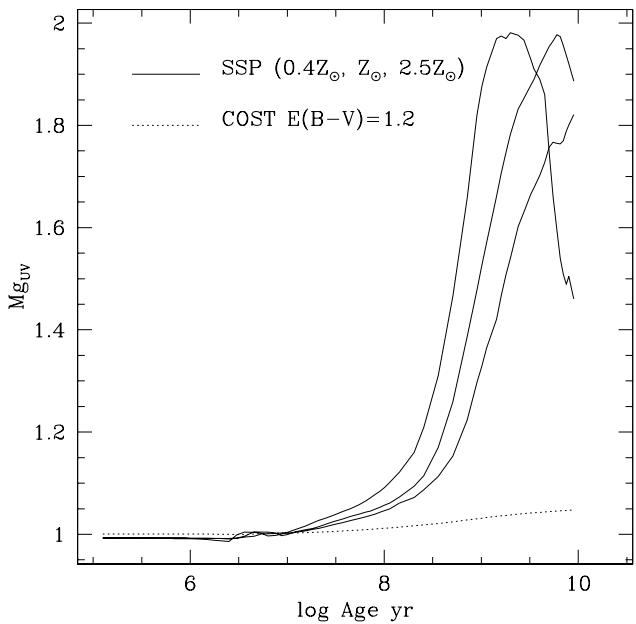}
\caption{The age dependence of the $Mg_{UV}$ index
(Eq.~\ref{eq:MgUV}). The three solid lines correspond to SSP models for
three different metallicities ($0.4Z_\odot, Z_\odot, 2.5Z_\odot$; 
higher metallicity produce a larger \mguv index at fixed age for most of the age range). 
The 
dotted line is for a constant star-formation model with $E(B-V)=1.2$.
The $Mg_{UV}$ index is almost independent on reddening and
on the resolution of the spectra for $R\simgt50$, the typical rest-frame 
resolution of our GRAPES spectra.
}
\label{fig:MgUV2}
\end{figure}

\section{UV spectra of passive galaxies: the M$\lowercase{g}_{UV}$ feature}
\label{sect:MgUV}

The low resolution of the
HST grism spectra does not allow us to detect individual
absorption or emission lines (if not for large equivalent
widths, Xu et al. 2005) whose identification 
generally yields redshift measurements in higher resolution
spectroscopy. Nevertheless the GRAPES spectra are well suited
for the identification of broad, low resolution spectral
features like the strong breaks that are commonly found
among old stellar populations. 
Fig.~\ref{fig:MgUV1} shows rest frame UV spectra 
(the region generally
accessible to GRAPES HST+ACS for $1.4<z<2.5$ objects) from Bruzual 
\& Charlot (2003) spectral synthesis models and from Kurucz models of stars
(Kurucz et al. 1979), 
smoothed to the typical resolution of our data.
While the 4000\AA\ break and the 3600\AA\ Balmer break rapidly disappear
beyond 1$\mu$m for $z\simgt1.4$, 
there are other strong age dependent
features in the UV that are detectable in relatively low resolution 
spectra. The most prominent one is a bump in the region
2640-2850\AA\ that
is due to the combination of several strong absorption lines,
including MgII2800 that is the
strongest one.
The typical shape of this region was first used by
Spinrad et al. (1996) to measure the redshift of the $z=1.55$ passive
galaxy, and more recently also by Cimatti et al. (2004) and McCarthy 
et al. (2004).
We have dubbed this feature the $Mg_{UV}$ feature, and defined the $Mg_{UV}$
index as:
\beq
Mg_{UV} = \frac{2\times\int_{2625}^{2725} f_\lambda
d\lambda}{\int_{2525}^{2625} f_\lambda
d\lambda+\int_{2725}^{2825}
f_\lambda d\lambda}
\label{eq:MgUV}
\eeq
where the integration ranges (see Fig.~\ref{fig:MgUV1}) are defined in \AA.
The \mguv feature is found to be almost independent of the spectral resolution
for $R\simgt50$, the typical rest-frame resolution of the spectra discussed 
in this paper, and on dust extinction.
Fig.~\ref{fig:MgUV2} shows the age dependence of the \mguv feature for 
passive as well as actively star-forming galaxies.
The $Mg_{UV}$ feature is a key fingerprint of the presence
of passively evolving stellar populations because it is not present in
young dust reddened star-forming galaxies
(Fig.~\ref{fig:MgUV1} and \ref{fig:MgUV2}), which can still produce 
similarly red overall colors for high dust extinction. 
Its secure detection therefore
allows one to establish the nature of the sources by
breaking the age/dusty degeneracy among red
sources, as well as the measurement of the galaxy redshift.

\section{Passive $z>1.4$ galaxies: sample selection}
\label{sect:BzK}

In order to search for high-redshift, passively evolving galaxies 
among the $\simgt10000$ 
galaxies detected in the UDF we relied on their peculiar color
properties.
Candidates of $z\simgt1.4$ passively evolving galaxies were pre-selected
following the criteria outlined by Daddi et al. (2004b). These were
calibrated on the complete K20 survey spectra database that includes
the $z>1.4$ passive and star-forming galaxies described in 
Cimatti et al. (2004) and Daddi et al. (2004a).
Defining the color difference:
$BzK\equiv (z-K)_{AB}-(B-z)_{AB}$, 
the candidate $z\sim2$ passive galaxies can be located with
$BzK<-0.2\ \ \ \bigcap\ \ \ (z-K)_{AB}>2.5$
(Daddi et al. 2004b).

Objects satisfying the above condition were retained as a primary
sample. However, based
on Fig.~8 (bottom-left diagram) of Daddi et al. (2004b), 
also galaxies having $z-K>2.5$ and $BzK>-0.2$ were considered 
for detailed analysis, 
as young "proto-ellipticals" could be in
principle located at $BzK>-0.2$. Moreover, despite the extreme
depth of $B$-band UDF imaging the errors on $B-z$ colors are
found to be large for many red objects with $z-K>2.5$.

\begin{figure}[ht]
\centering
\includegraphics[width=8.8cm]{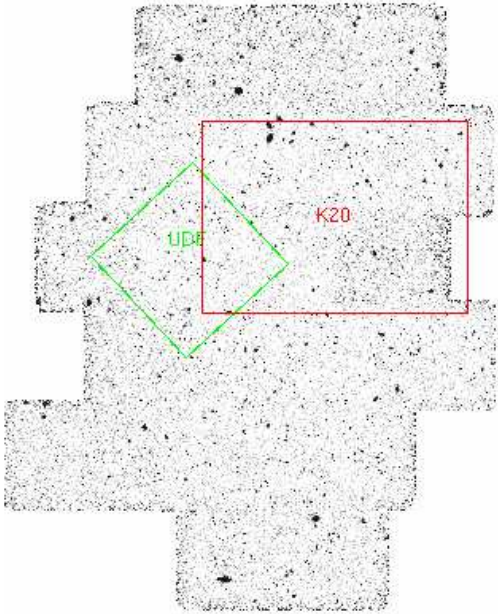}
\caption{The relative layout of the UDF, K20 and GOODS-South
regions. The background image is the GOODS K-band mosaic used in this
paper. 
}
\label{fig:reg_UDF}
\end{figure}

This pre-selection criteria require $B$-, $z$-, and $K$-band imaging. 
We have used the deep VLT+ISAAC $Ks$-band images\footnote{Available at
"http://www.eso.org/science/goods/releases/20040430/"} over the GOODS
field (Fig.~\ref{fig:reg_UDF})
(Vandame et al. 2005 in preparation), 
and the ultra-deep ACS images\footnote{Available at
"http://archive.stsci.edu/prepds/udf/udf\_hlsp.html"} of the UDF for
the $B-$ and $z-$bands (Beckwith et al. 2005 in preparation). 
Galaxies were selected
over the area of about 12 arcmin$^2$ where 
deep ACS images are available, requiring total magnitudes
$K_{AB}<23$ as measured by SExtractor MAG\_AUTO (Bertin \& Arnouts
1996).
This is $\sim1$ mag fainter than reached by the K20 survey, while the
area is a factor $\sim3$ smaller than the K20/GOODS area
that is contiguous and in part overlapping with the UDF area analysed here
(Fig.~\ref{fig:reg_UDF}).

Fig.~\ref{fig:BzK} show the resulting $BzK$ diagram. In order to
increase the accuracy of the color measurements we have not attempted to 
reduce the ACS resolution to the much poorer 0.5$''$ seeing of the
K-band data. For the ACS bands we used SExtractor MAG\_AUTO measured in
consistent apertures defined in the $z$-band (double image mode). 
SExtractor MAG\_AUTO measurements were used also for the K-band 
(single image mode). This
results in a different physical aperture of the magnitudes in the optical
versus near-IR bands. However, we verified that the resulting
$z-K$ was not biased (at least within $\sim0.05$ mags on average)  with
respect to seeing matched $z-K$ magnitudes for objects in common to the
catalog of Daddi et al. (2004b). 
There are 20 objects with $z-K>2.5$ of which 7 have also $BzK<-0.2$.

\section{Analysis: redshifts and spectra}
\label{sect:reds}

To measure the redshifts of the selected candidates, 
we have used all the  
available information including high quality HST low resolution optical 
spectra and the multicolor photometric data available for the UDF,
as discussed in the next two sections.

\subsection{ACS-HST low-resolution spectroscopy}

The ACS-HST grism 800L data, obtained as part of the GRAPES project,
cover the wavelength
range 5000-11000\AA\, with maximum efficiency around
7000-8000\AA. The spectral dispersion is about 40\AA/pixel. The low
sky background at these wavelengths from space allow HST grism data to 
be taken 
in slitless mode, so that all UDF objects are simultaneously observed 
in the GRAPES data.
The effective spectral resolution depends on the spatial
size of the objects. For unresolved sources the spectral dispersion 
translates into a resolution of $R\approx100$ at 8000\AA,
while extended sources have lower resolution spectra.
Thanks to the absence of OH sky emission lines and the exquisite spatial
resolution of HST, the GRAPES
low-resolution spectroscopy allows us to reach relatively high S/N 
ratios on the continuum. 
For compact galaxies, like the ones considered in this
paper, one gets $S/N\sim10$ at 8000\AA\ over a 100\AA\ region  for
$i_{AB}\sim25.5$, for the 10$^5$s integration time of the GRAPES data. 
This is significantly better than what can be reached
from the ground with
red-optimized CCDs on 8-10m class telescopes with similar
integration times, except for the narrow spectral windows free of OH  sky
emission lines. 

\begin{figure}[ht]
\centering 
\includegraphics[width=8.8cm]{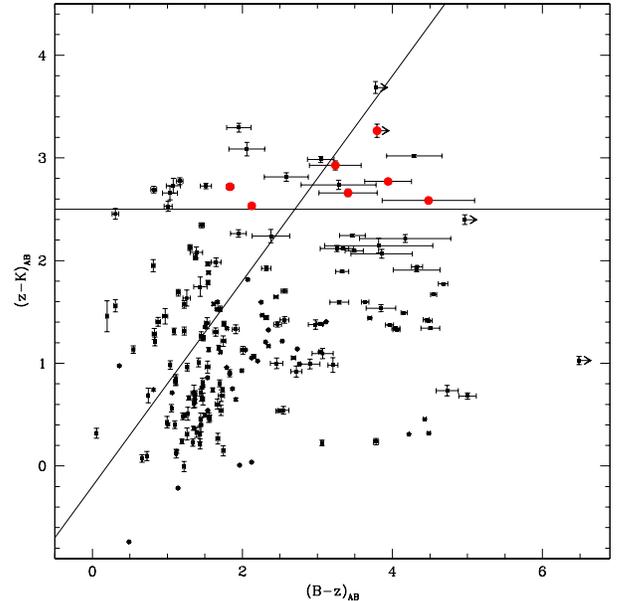}
\caption{The $BzK$ diagram for galaxies with $K_{AB}<23$ in the UDF. 
Objects with $z-K>2.5$ were retained as possible high-$z$ old
galaxies candidates. The seven
passive early-type galaxies with proposed redshift $1.4\simlt z\simlt2.5$ are
shown with large symbols. Five occupy the region with $BzK<-0.2$ as
expected (Daddi et al. 2004b), while two bluer objects are also
identified.
}
\label{fig:BzK}
\end{figure}

Full details of GRAPES grism data reduction and calibration  
are given in P04. As slitless spectroscopy can imply 
spectra superposition from neighboring sources, the data were obtained at 
five independent position angles to minimize this effect. 
Narrow extraction windows (see P04) were used to maximize 
the S/N ratio of the resulting spectra. The 5 epoch spectra were
co-added, avoiding regions contaminated by neighboring objects, as
described in P04.

\begin{deluxetable*}{ccccccccccccccc}
\tabletypesize{\tiny}
\tablecaption{Photometric properties, SED and spectral fitting results}
\tablewidth{0pt}
\tablehead{
\multicolumn{3}{c}{} & \multicolumn{6}{c}{Photometry} &
\multicolumn{3}{c}{SED fitting} & \multicolumn{3}{c}{Spectral fitting}\\
\colhead{ID} & \colhead{RA} & \colhead{DEC} & \colhead{$z^{tot}$} &
\colhead{$K^{tot}$} & \colhead{$B-z$} & \colhead{$z-K$} &
\colhead{$R-K$} & \colhead{$J-K$} & \colhead{$z_{phot}$} &
\colhead{$\chi^2_{\rm r,old}$} & \colhead{$\chi^2_{\rm r,d}$}
& \colhead{$z_{spec}$} &
\colhead{$\chi^2_{\rm r,old}$} & \colhead{$\chi^2_{\rm r,d}$}\\
\colhead{\ } & \colhead{\ } & \colhead{\ } & \colhead{AB} &
\colhead{AB} & \colhead{AB} & \colhead{AB} & \colhead{Vega} & 
\colhead{Vega} & \colhead{\ } &\colhead{\ } & \colhead{\ }
& \colhead{\ } &\colhead{\ } & \colhead{\ }
}
\startdata
\label{tab:data}
8238 & 03:32:36.9 & -27:46:28.5 & 23.80 & 21.22 & 4.48 & 2.58 & 6.43 & 2.01
& $1.26\pm0.04$ & 1.1 & 2.8 & $1.39\pm0.01$ & 0.59 & 0.61\\ 
4950 & 03:32:30.0 & -27:47:26.8 & 22.56 & 19.90 & 2.12 & 2.64 & 5.58 & 1.96
& $1.57\pm0.03$ & 0.4 & 7.2 & $1.55\pm0.01$ & 0.83 & 0.79\\ 
1025 & 03:32:43.0 & -27:48:45.1 & 24.20 & 21.48 & 1.83 & 2.72 & 5.41 & 1.87
& $1.72\pm0.04$ & 0.6 & 5.1 & $1.73\pm0.01$ & 0.40 & 0.54\\ 
3523 & 03:32:33.7 & -27:47:51.1 & 24.34 & 21.68 & 3.41 & 2.66 & 5.84 & 1.79
& $1.72\pm0.04$ & 2.2 & 9.0 & $1.76\pm0.02$ & 0.70$^a$ & 0.64\\ 
3650 & 03:32:38.1 & -27:47:49.8 & 24.18 & 21.36 & 3.94 & 2.82 & 6.00 & 1.84
& $1.90\pm0.03$ & 2.2 & 9.7 & $1.91\pm0.01$ & 0.76 & 0.98\\ 
3574 & 03:32:39.1 & -27:47:51.6 & 25.29 & 22.39 & 3.24 & 2.90 & 6.29 & 1.89
& $1.94\pm0.04$ & 3.0 & 11.3 & $1.98\pm0.02$ & 0.41 & 0.46\\ 
1446 & 03:32:39.2 & -27:48:32.4 & 26.06 & 22.80 & $>$3.80 & 3.26 & 5.42 & 3.32
& $2.85\pm0.05$ & 2.1 & 5.8 & $2.47\pm0.02$ & 0.58 & 0.71
\enddata
\tablecomments{IDs for UDF galaxies are those from the publicly
available catalog release V1, available at
"ftp://udf.eso.org/archive/pub/udf/acs-wfc/h\_udf\_wfc\_V1\_cat.txt". 
Note that the total magnitudes were
derived from S\'ersic profile fitting, as described in
Sect.~\ref{sec:morpho}, and are brighter than the Kron total magnitude 
measured by SExtractor by $\sim0.4$ mags, on average.
Conversion factors from Vega to AB magnitudes
are (-0.09, 0.22, 0.94, 1.87) for ($B$, $R$, $J$ and $K$), respectively.
Photometric redshift errors are at the 68\% level as
computed by {\em hyperz}. Spectroscopic redshift errors are the formal 68\%
range on the fit determined following Avni (1976) for the case of one
interesting parameter (redshift), added in quadrature to
systematic and random errors in the wavelength calibration that are about
20\AA\ (Pasquali et al. 2003; P04) and
produce an additional uncertainty of $\sigma_z\simlt0.01$.
$^a$ The spectral fit for object \#3523
has a slightly lower minimum for $z\sim1.1$, a redshift that is however
fully inconsistent with the photometric SED.
}
\end{deluxetable*}

\subsection{Spectral analysis}

\begin{figure*}[ht]
\centering 
\includegraphics[width=18cm]{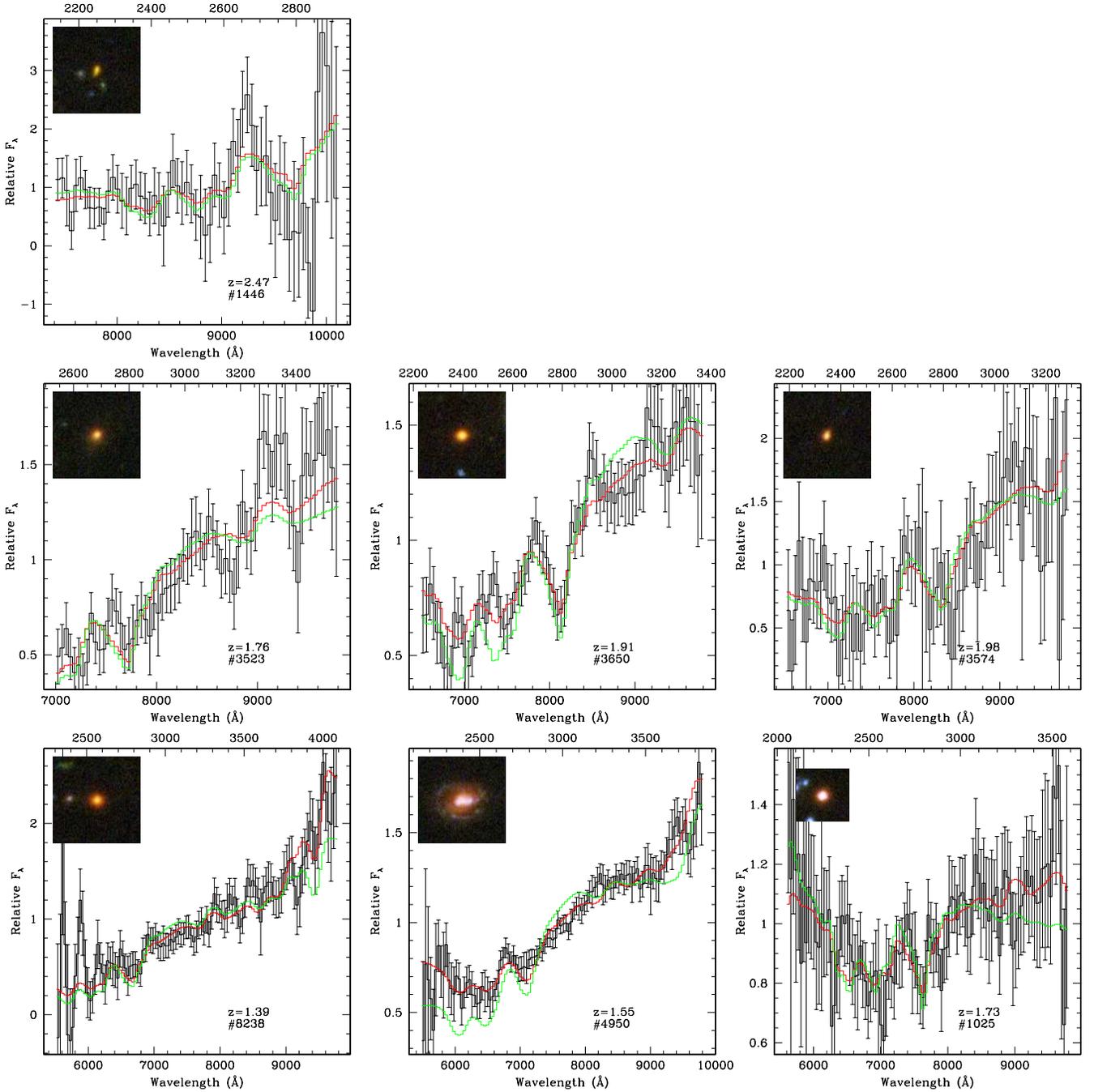}
\caption{The GRAPES HST+ACS spectra of the seven passively evolving galaxies
with $1.39<z<2.47$. In each panel the observed spectra are fitted with
Bruzual \& Charlot 
(2003) models for old galaxies (red) and main sequence stars from the Kurucz
library (green). Proposed redshifts and UDF IDs are labeled on each
panel. Both rest-frame (top x-axis) and observed (bottom x-axis)
wavelengths are labeled.
Color inserts for each galaxies are $5''\times5''$ (except for \#1025
that has a smaller color image because close to the edge of the UDF
field).
}
\label{fig:speALL}
\end{figure*}

The 20 red galaxies selected as described in Sect.~\ref{sect:BzK} were
inspected by looking for the \mguv feature described above, and to their
overall spectral shape. Some of the spectra had too low S/N ratio 
to be useful because the objects are too faint in the optical.
Seven objects were retained as likely $z\simgt1.4$ passive objects
identifications, five of these having $BzK<-0.2$. 
Spectra for these seven galaxies are shown in
Fig.~\ref{fig:speALL}. The five single epoch spectra of each object
were inspected to verify that the detected features were persistent
and not due to obvious spurious artifacts.

In order to objectively identify the redshift of the objects, and to
exclude possible features misidentification,
we cross correlated the spectra with galaxy templates from
the Bruzual \& Charlot  (2003) library before definitively
assigning redshifts. The model templates 
were convolved with the line spread functions computed from
the light profiles of the galaxies to reproduce the actual
spectral resolution of the data. 
Note that this fitting approach is rather conservative because most of the
continuum region of the spectra with no detected
features have no strong effect on the fitting 
and reduce the impact of the regions where the actual features are,
i.e. it dilutes the signal by giving equal weight to all data in the
spectral range.

Two classes of models were adopted for the spectral fits in order
to bracket the general  cases of old and passive galaxies versus 
strongly dust reddened galaxies. For old galaxies we considered
simple stellar populations (SSP) and a model with exponentially
declining SFR with timescale $\tau=0.3$ Gyr. Ages less than the age of
the universe at each redshift were required and a maximum reddening of
$E(B-V)<0.2$ was allowed. We considered solar and 40\% solar 
metallicity. For dusty galaxies we used
constant star-formation rate models with unlimited\footnote{practically, for the intrinsically bluest models,
$E(B-V)\simlt1.5$ is more than sufficient to provide SEDs as red or redder than the data at any redshifts} dust reddening 
and solar metallicity.

In all seven cases, except for \#3523, the models 
for old passive galaxies converge to the solution
guessed by eye, with the \mguv feature correctly identified. 
In most cases dusty models provide a worse 
fit to the data, however, differences are not large and all reduced
$\chi^2$ are below 1 and acceptable. The minima of reduced $\chi^2$
being all below 1 might actually suggest that some significant degree of 
correlation (e.g., due to the reduction process, P04) is present in the data.
Results of the spectral fitting are summarized in Table~\ref{tab:data},
where in all cases the $z_{spec}$ is quoted for the best fitting
old/passive galaxy model, as justified from the results of the next
section.
Detailed description of redshift assignment for
individual objects is described in Sect.~\ref{sec:assi}, following 
the analysis of the SEDs and photometric redshifts.

\begin{figure*}[ht]
\centering 
\includegraphics[width=13cm,angle=-90]{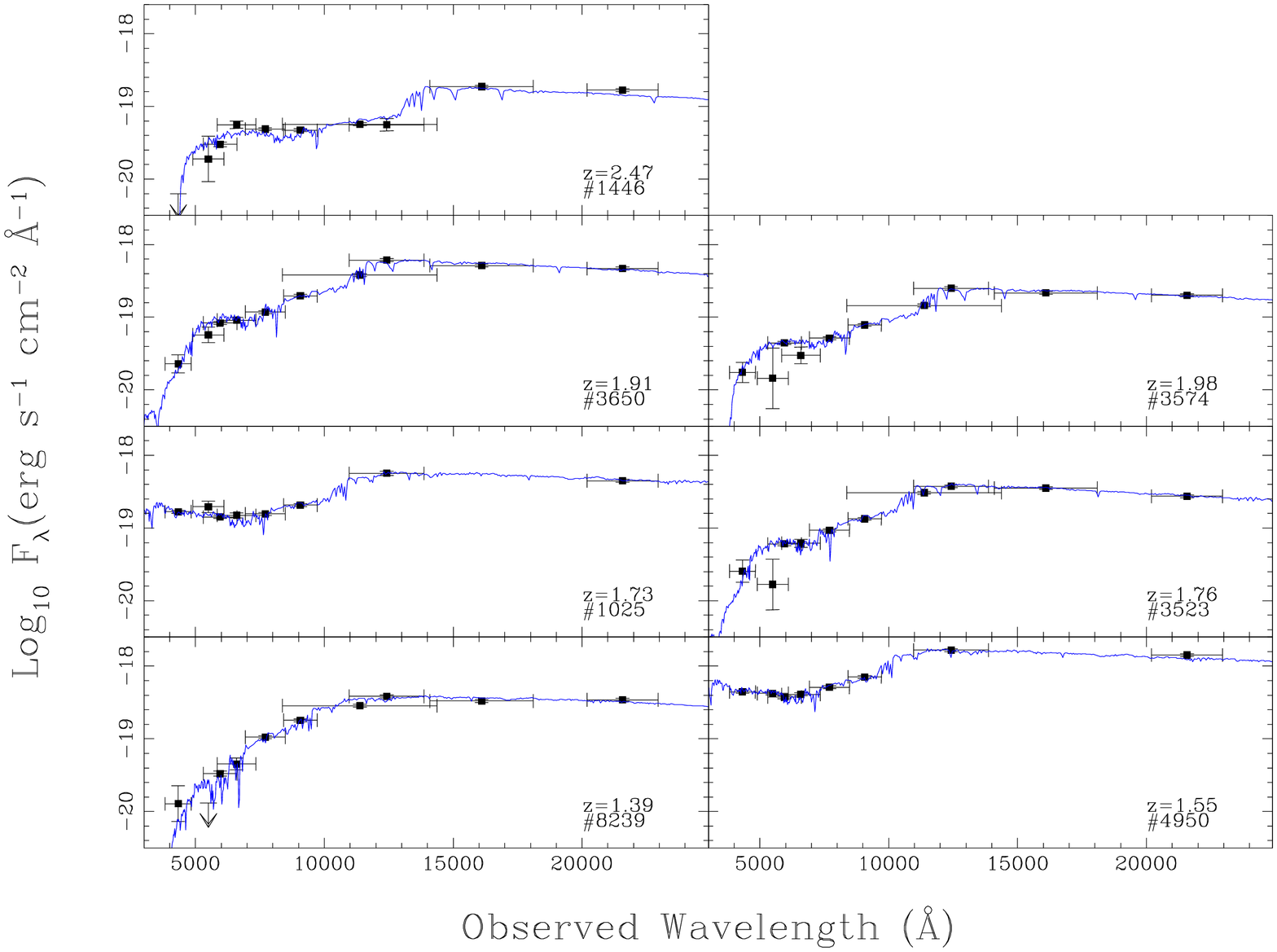}
\caption{The observed spectral energy distributions of the seven,
$z\simgt1.4$ 
early-type galaxies. The imaging data shown in the plot are, from blue to
red: F435W, V, F606W, R, F775W, F850LP, F110W, J, F160W, K. The
best fit Bruzual \& Charlot (2003) SSP models overplotted here are
consistent with those
used in Fig.~5. Redshift is
increasing from bottom to top. The upper limits shown are at 1$\sigma$.
}
\label{fig:SEDs}
\end{figure*}

\subsection{Spectral energy distributions and photometric redshifts}
\label{sec:photoz}

To further constrain the redshift identifications we used the
extremely high-quality imaging available on the UDF area. This includes
the $BViz$ ACS imaging 
and the J and H NICMOS imaging. ISAAC J and K band data were also used,
the J-band filter of ISAAC being much narrower than the
NICMOS one,
as well as V and R from FORS2 that have central wavelengths that are 
quite different from the ones of
the ACS bands.  To match ground based and HST datasets we proceeded as
discussed in Sect.~\ref{sect:BzK}.
It was checked that over the common range the photometric
SEDs and the HST low resolution spectra were consistent. The SED of the
seven $z>1.4$ early-type galaxies candidates are shown if
Fig.~\ref{fig:SEDs}.
The same range of models parameters as for the
spectral fitting was used, corresponding to the two classes of
old/passive versus dusty models. The {\em hyperz} code
(Bolzonella et al. 2000) was used for the model fitting and to derive
photometric redshifts. A lower limit of 0.05 mags was used for the 
photometric errors in
all the bands, in order to account for photometric zeropoint uncertainties
and possible residual uncertainties in the match of the different datasets. 

Results of the SED fitting and photometric measurements
are summarized in Table~\ref{tab:data}.
SED fitting in all cases result in significantly better fits for models
of old/passive galaxies with respect to those for dusty objects,
rejecting dusty models at more than $3\sigma$ confidence levels 
(Avni 1976) in all cases. 
Best fits have generally acceptable
reduced $\chi^2$ and indeed look reasonable by eye inspection.
The agreement between the photometric redshifts and the
spectroscopic redshifts, derived from the spectral fitting with old stellar
populations, is very good with 6 out of 7 objects having
$|z_{spec}-z_{phot}|\simlt0.1$.

\subsection{Redshifts and redshifts quality classes for individual objects}
\label{sec:assi}

Based on the results of SED and spectral fitting, we assigned redshifts
and redshifts quality classes ($A$ for good-confidence redshifts, $B$
for less secure ones) for individual objects. A quality "A" redshift 
is assigned when the \mguv feature appears well detected, and when the
photometric and spectroscopic redshifts are in good agreement (i.e.,
within $\Delta z=0.1$).

\spb \#8238: $z=1.39$, Class $=$ B.\\
The SED and spectral fitting agree within $\Delta z=0.13$, the spectrum
is consistent with the presence of the 4000\AA\ break and of the \mguv
feature.

\spb \#4950: $z=1.55$, Class $=$ A.\\
The SED and spectral fitting agree very closely with $\Delta z=0.02$, 
the spectrum clearly shows the \mguv feature. The rise
toward the UV below 6000\AA\ is consistently present also in the photometry.
This object is also part of the K20 survey (Cimatti et al. 2002b)
with a measured redshift of $z=1.553$. A weak [OII]$\lambda$3727
emission line is detected in the K20 survey spectrum.

\spb \#1025: $z=1.73$, Class $=$ A.\\
The SED and spectral fitting agree extremely closely with $\Delta
z=0.01$. The spectrum, that is fairly blue especially at
$\lambda<2500$\AA,
shows the characteristic \mguv feature. The upturn in the blue spectra
is consistent with what expected for A-type stars
(Fig.~\ref{fig:MgUV1}).
This source is detected in the X-rays (Giacconi et al. 2002; Alexander
et al. 2003; see Sect.~\ref{sec:Xrays}).
Zheng et al. (2004) propose $1.47<z_{phot}<1.58$,
reasonably consistent with our spectroscopic redshift.

\spb \#3523: $z=1.76$, Class $=$ B.\\
The spectrum would suggest $z\sim1.1$ as the best fitting
old galaxy solution by placing
the 4000\AA\ break around 9000\AA\
and the CaII H\&K lines in the dip at 8700\AA. The above
solution is not consistent with the
photometric SED, as at $z\approx 1.1$ no 
reasonably good fit could be obtained.
A secondary solution from spectral fitting,
with nearly the same $\chi^2$ of the best one,
is found at $z=1.76$, in very good agreement with the photometric
redshift within $\Delta z=0.04$. We propose the latter solution
as the spectroscopic redshift for this galaxy.
The \mguv feature is not evident, although the spectrum is consistent at
least with the presence of a 2900\AA\ break. The proposed redshift
thus relies strongly on the overall SED and spectral fitting. The absorption
like spectral features at 8900\AA\ and 9200\AA\ are to be considered spurious
noise features for our proposed redshift.
Yan et al. (2004) suggest this is an old galaxy at $z_{phot}=1.6$, in 
good agreement with the redshift proposed here.

\spb \#3650: $z=1.91$, Class $=$ A.\\
The SED and spectral fitting agree extremely closely with $\Delta
z=0.01$. The \mguv feature is very evident.

\spb \#3574: $z=1.98$, Class $=$ A.\\
The SED and spectral fitting agree closely with $\Delta z=0.04$. 
The spectrum shows the \mguv feature.
Yan et al.
(2004) propose this is an old galaxy at $z_{phot}=1.9$, in excellent
agreement with our spectroscopic redshift.

\spb \#1446: $z=2.47$, Class $=$ B.\\
The SED and spectral fitting are in fair agreement at best
with $\Delta z\sim0.4$. The spectrum has a strong
feature at 9200\AA\ that in the spectral fitting with old models is
identified with the \mguv feature. Emission line identification algorithms
described in Xu et al. (2005), however, pick up the feature as a possible
emission line. If the feature is an emission line we find best fitting
$\lambda\sim9230$\AA, FWHM$\sim160$\AA\ and EW$\sim150$\AA.
The large FWHM and EW would suggest an AGN emission line.
This would be possible as the object is an hard X-ray source (Giacconi
et al. 2002; Alexander et al. 2004), although no strong emission line is 
detected in the FORS spectra described by Szokoly et al. (2004).
We tried to fit its photometric SED
with reddened QSO/AGN templates but the fit still significantly prefers 
old/passive models. 
As the SED drops in the B-band, the only plausible AGN line
identification would be [CIII]$\lambda$1909 at $z=3.83$, with the Lyman-break
producing the B-band drop. 
Zheng et al. (2004) propose
a photometric redshift $4.13<z_{phot}<4.35$, not far from that.
For $z=3.83$ one would expect to detect an even stronger
CIV$\lambda$1550 emission at about 7500\AA\ that instead is not present.
For $z=3.83$ it would be difficult also
to explain the overall shape of the SED with,
e.g., the break between the J- and H-bands. 
Chen \& Marzke (2004) propose this object is a dusty galaxy with
$z_{phot}=3.43$, not consistently with our analysis, while 
Yan et al.
(2004) propose this is an old galaxy at $z_{phot}=2.8$.
\bigskip

The spectral and photometric properties show that we can distinguish two
classes of objects among our sources. Objects \#4950 and \#1025 have in
fact significant B-band excess flux with respect to the others (see
Fig.~\ref{fig:BzK} and Fig.~\ref{fig:images}) suggesting the presence
of hotter stars. These two objects will be discussed
in more details in Sect.~\ref{sec:blues}.

\begin{figure}[ht]
\centering 
\includegraphics[width=8.8cm]{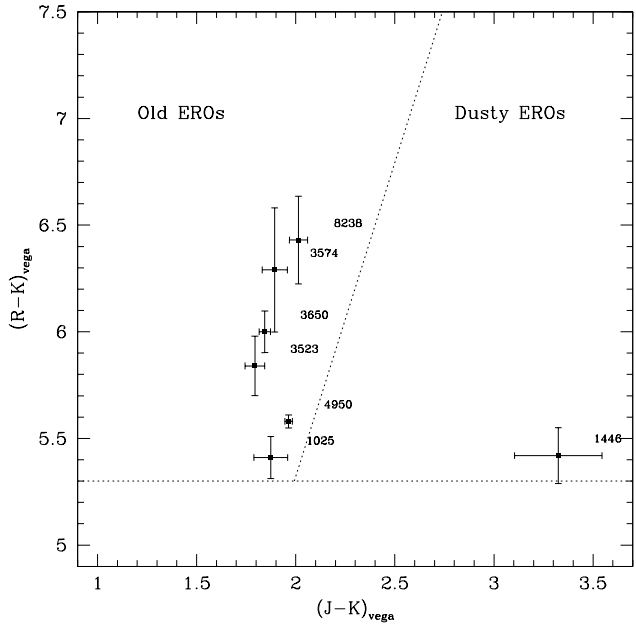}
\caption{The Pozzetti \& Mannucci (2000) diagnostic diagram with the 
$R-K$ versus $J-K$ colors of our sample galaxies. The dotted lines show
the $R-K>5.3$ limit of validity of the criterion and the division
between old and dusty star-forming EROs for $z\simlt2$.
}
\label{fig:PM00}
\end{figure}

\subsection{Discussion of redshift identifications}

The proposed redshifts are based on low resolution spectra that, we 
recall, do not allow
to detect individual absorption or emission lines as ordinarily
done for redshift identifications of faint galaxy spectroscopy.
Therefore, we cannot exclude with full confidence 
that, in a minority of the cases, our identifications might be somehow
mistaken.
We believe, however, that identifications are generally reliable,
because: (1) the spectral fitting routine generally converge to
the solution guessed by visual inspection of the spectra, when using
templates for passive galaxies; (2) the
photometric redshift estimated with the high-quality UDF data are in
very good agreement with the proposed spectroscopic redshifts with 
median $|z_{spec}-z_{phot}|=0.04$ only; 
(3) there is an overall consistent picture as signatures of old 
stellar populations are seen in the spectra, the
SEDs strongly argue for passive stellar populations and morphologies of 
early-type galaxies are recovered, as discussed in the next section.

As an additional test, we verified the location of the seven
sources in the Pozzetti \& Mannucci (2000) diagram, Fig.~\ref{fig:PM00}. 
The $R-J$ versus $J-K$ colors of the sources are consistent with those
expected for passively evolving galaxies up to $z\simlt2$. The only due
exception is object \#1446 with much redder $J-K$ color and proposed 
$z=2.47$, a redshift at which the above diagnostic may not apply.
We verified that using the criterion proposed by Pozzetti \& Mannucci (2000) 
for $z>2$, involving the J-, H- and K- bands, also the object \#1446 would
be classified as a passive red galaxy.

We also verified, for further overall consistency,
that best fitting models of the SEDs produce good fits
to the HST grism spectra.

\subsection{Sample completeness}
\label{sec:compl}

The completeness of the $z\simgt 1.4$ sample of passive
galaxies was investigated. We
checked if additional passive $z\simgt 1.4$ galaxies could be found
among bluer objects with $2.2<z-K<2.5$. It may be expected that toward
the lowest $z\sim1.4$ range of redshifts some early type galaxy could
have 
$z-K<2.5$ (either intrinsically or because of the errors in the photometry). 
No convincing case could be found.
Among the remaining unidentified galaxies with $z-K>2.5$
there are 3 more objects with compact 
morphology that are not in our sample with proposed spectroscopic
redshift. One of these is \#869, which
is a $z=3.064$ type 2 QSO
(Szokoly et al. 2004), with no distinctive feature in our
GRAPES spectrum.
The other two are UDF \#8363 and \#5056, and have noisy
spectra with no feature identifiable and photometric redshifts 2.30 and
1.43, respectively. These might be additional passive $z>1.4$ objects.
Their $BzK$ colors 
are consistent with the $BzK<-0.2$ region (object \#8363 is the
reddest one in $z-K$ with a lower limit to  $B-z$, Fig.~\ref{fig:BzK}).

There are 10 more galaxies with $z-K>2.5$ that remain unidentified
in GRAPES spectra, and
have irregular, diffuse and sometimes merging-like morphologies,
reminiscent of the $z\sim2$ star-forming galaxies (having also similar colors 
and magnitudes) shown in Daddi et al. (2004a,b). 
Only one of these has formally $BzK<-0.2$.
The fact that we cannot detect features of old stars in these objects could 
be an observational bias as the S/N of the spectra for these extended
and diffuse galaxies is generally low. However, they might be more likely
genuine dust reddened star-forming galaxies. An analysis of the stacked 
X-ray and radio emission of galaxies with similarly $z-K$ colors and having
$BzK<-0.2$ in the K20 survey suggests that these objects are mainly
vigorous dust reddened starbursts at $z>1.4$ (Daddi et al. 2004a,b).

In summary, there are at most only two reasonable additional 
candidates passive $z>1.4$ galaxies, that are compact
and have the expected $BzK$ colors. These are less luminous than the ones
for which we propose a spectroscopic redshift. Additional
passive objects might still be present among the galaxies
with irregular morphology, although no object with similar
properties is currently known both at high and lower
redshift.

\subsection{Redshift clustering}

We mention in passing the few evidences of redshift clustering
that can be drawn from the redshifts of our sample.
For these objects, the correlation length $r_0$ might be comparable or even 
higher than for typical EROs at $z\sim1$ ($r_0\sim10$ \h1 Mpc; Daddi et al.
2001; McCarthy et al. 2001; Brown et al. 2005), 
as these could be the truly first 
collapsed massive galaxies. This should translate in significant
pairings
between ours and to other galaxies, e.g. to those in the K20 region
(Fig.~\ref{fig:reg_UDF}). In fact, the $z=1.91$ galaxy is at the same
redshift of the Cimatti et al. (2004) $z=1.903$ elliptical and also to
that of the $z=1.901$ near-IR bright galaxy in Daddi et al. (2004b).
The $z=1.73$ and $z=1.76$ galaxies in our sample may actually be at the
same redshift given the errors, presumably $z=1.73$ where two other
near-IR bright massive star-forming galaxies also lie
(Daddi et al. 2004b). Also at $z=1.55$
and $z=1.39$ other massive star-forming galaxies are found in the K20
survey. Similarly, 3 out of 4 of the $z>1.5$ early-type galaxies
by Cimatti et al. (2004) are at $z\sim1.61$, and strong pairing is
observed also in the McCarthy et al. (2004) sample.

\begin{figure*}[ht]
\centering 
\includegraphics[width=17cm]{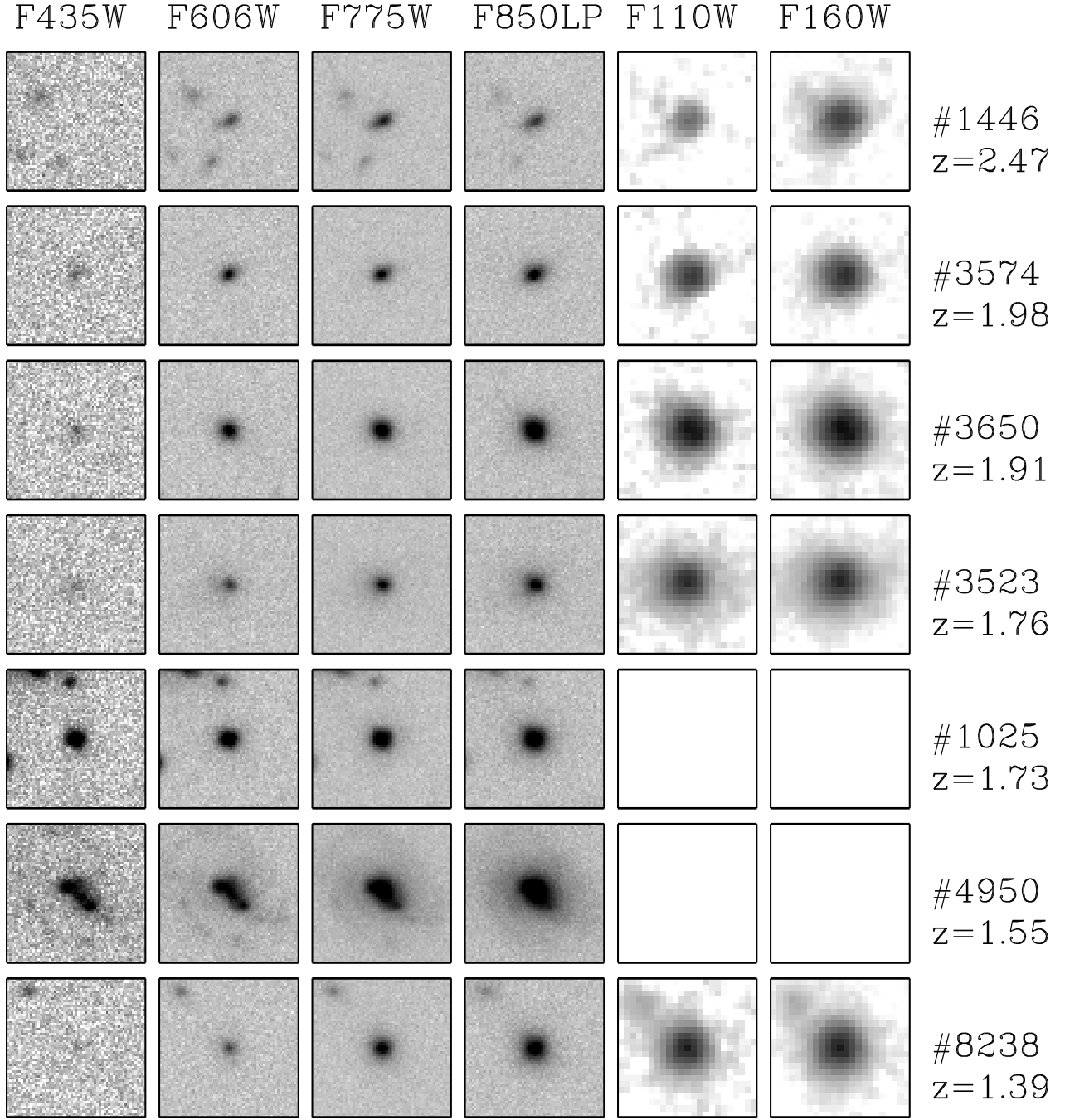}
\caption{Multiband HST imaging of the high-$z$ early-type galaxies.
Each image is 2$''$ on a side.
Within the ACS imaging, as within the NICMOS imaging but on a different
scale, the same range of
fluxes ($f_\lambda$) are plotted in a logarithmic scale, to provide
direct comparison with Fig.~\ref{fig:SEDs} and among different objects
or different bands.
The F435W band appears the noisiest in $f_\lambda$,
where however only object \#1446 is not significantly detected. 
The NICMOS images allow in some cases to see the
extended, low-surface brightness wings of the steep de Vauculeurs like
profiles. Objects \#1025 and \#4950 are outside of the region covered by
the NICMOS imaging.
}
\label{fig:images}
\end{figure*}

\begin{deluxetable*}{rccccccccccccc}
\tabletypesize{\scriptsize}
\tablecaption{Morphological parameters}
\tablewidth{0pt}
\tablehead{
\multicolumn{1}{r}{} & \multicolumn{1}{c}{} &  \multicolumn{2}{c}{$i$-band} & \multicolumn{5}{c}{$z$-band} & \multicolumn{5}{c}{$i$-band}\\
\colhead{ID} & \colhead{Type} & \colhead{C} & \colhead{A} 
& \colhead{n} & \colhead{$r_e$ pix} & \colhead{$r_e$ kpc} & \colhead{b/a}
& \colhead{n} & \colhead{$r_e$ pix} & \colhead{$r_e$ kpc} & \colhead{b/a}
}
\startdata
\label{tab:galfit}
8238 & E/S0 & 3.09 & 0.11 & $8.2\pm1.5$ & $11\pm4.7$ & $2.8\pm1.2$ &
0.89 &
$5.8\pm1.0$ & $4.7\pm1.1$ & $1.2\pm0.3$ & 0.88 \\
4950 & Sa+M? & 3.57 & 0.18 & $4.3\pm0.4$ & $22\pm3.7$ & $5.6\pm0.9$ &
0.60 & -- & -- & -- & -- \\
1025 & E/S0 & 3.59 & 0.12 & $4.2\pm0.5$ & $2.9\pm0.3$ & $0.74\pm0.1$ &
0.85 &
$5.0\pm0.7$ & $2.1\pm0.3$ & $0.54\pm0.1$ & 0.85 \\
3523 & E/S0 & 4.16 & 0.19 & $9.0\pm1.6$ & $11\pm4.3$ & $2.8\pm1.1$ &
0.74 &
$9.7\pm2.0$ & $16\pm8.0$  & $4.0\pm2.0$ & 0.71 \\
3650 & E/S0 & 2.87 & 0.12 & $4.7\pm0.6$ & $3.1\pm0.3$ & $0.79\pm0.08$ &
0.74 &
$5.4\pm0.8$ & $3.2\pm0.5$  & $0.81\pm0.13$ & 0.72 \\
3574 & E/S0 & 2.72 & 0.12 & $2.9\pm0.3$ & $2.5\pm0.2$ & $0.63\pm0.05$ &
0.45 &
$2.8\pm0.4$ & $2.4\pm0.4$  & $0.61\pm0.1$ & 0.39  \\
1446 & S0/Sa & 2.81 & 0.16 & $0.8\pm0.2$ & $3.1\pm0.5$ & $0.76\pm0.12$ &
0.35 &
$1.4\pm0.3$ & $3.9\pm0.8$  & $0.96\pm0.20$ & 0.26
\enddata
\end{deluxetable*}

\section{Analysis: morphology}
\label{sec:galfit}

We inspected by eye the morphological appearance of the 7 galaxies with
proposed spectroscopic confirmation as old high-z stellar populations.
Fig.~\ref{fig:images} show the available HST imaging (see also the color
images of the objects in Fig.~\ref{fig:speALL}).
All objects look
very compact and regular. Object \#4950
appears a bulge dominated nearly face-on spiral, i.e. an {\em Sa},
with perhaps evidence of merging or cannibalism in the blue bands.
Object \#1446 is the most elongated one and might be an early spiral 
or an S0 galaxy.
The other objects are visually classified as E/S0 systems.
In order to provide a more quantitative estimate of the morphology we
used CAS and S\'ersic profile fitting.

\subsection{CAS analysis}

\begin{figure}[ht]
\centering 
\includegraphics[width=8.8cm]{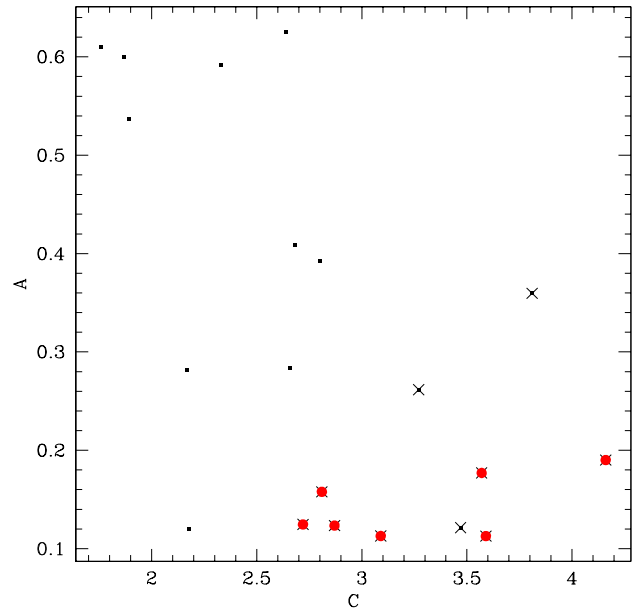}
\caption{The concentration versus asymmetry parameters in the $i$-band for
the 20 galaxies with $z-K>2.5$ and $K_{AB}<23$ in the UDF. Large circles
are for the seven $z\simgt1.4$ early-type galaxies. 
A cross around a symbol
is for objects classified as "compact" by visual inspection, as opposed to the
remaining objects (shown without crosses) that are classified as "irregular/merger".
The three compact objects that are not in our list of passively evolving
galaxies are described in more detail in Sect.~\ref{sec:compl}.
}
\label{fig:CAS}
\end{figure}

The concentration and asymmetry of all the galaxies
having $z-K>2.5$
were measured
in the $i$-band, corresponding to 
about 3000\AA\ rest-frame for $z\sim2$ (similar results would be
obtained in the $z$-band). 
Parameters definitions
consistent with those given in Conselice (2003) were used.
Fig.~\ref{fig:CAS} shows the resulting measurements.
Objects with proposed spectroscopic confirmation as $z>1.4$ passive
galaxies tend to
have high concentrations and low asymmetry (see also
Tab.~\ref{tab:galfit}).
In general, these objects occupy a range of values consistent with those
of early-type galaxies: for example all have concentrations $C>2.6$, 
a value marking the boundary between early- and late-type galaxies 
in the local universe (Strateva et al. 2001; see also Conselice 
2003). The asymmetries 
are generally low: $\simlt0.2$ in all cases.

\subsection{S\'ersic profile fitting}
\label{sec:morpho}

We used GALFIT (Peng et al. 2002)
to model the light profile of the 7 objects with
proposed redshift confirmation. Both $i$-band and $z$-band analyses
were performed for a cross check of the results and to test for possible
color trends. The PSF was derived for each band by averaging stellar
objects in the field (Pirzkal et al. 2005). 
All the $z\simgt1.4$ passive galaxies are resolved. 
We fitted S\'ersic (1968) profiles of the form:
\beq
\mu(r) = \mu_e \mbox{ e}^{-\kappa \left[({\frac{r}{r_e}})^{1/n} - 1\right]}
\eeq
where $r_e$ is the effective radius, $\mu_e$ is the effective surface
brightness, $n$ is the free S\'ersic index, and $\kappa$ is determined from
$n$ in order for half of the integrated flux to be within $r_e$.
Results are summarized in Table~\ref{tab:galfit}. 
The S\'ersic profile fitting also allowed us to derive the total
magnitudes, as listed in Table~\ref{tab:data}. 
The quoted uncertainties are purely statistical and derived by 
GALFIT on 
the basis 
of the image noise. Notice that the errors on the parameters are
correlated. For example, fitting with a larger $n$ generally implies
larger $r_e$ and brighter 
total magnitudes.

The S\'ersic index $n$ is generally found consistent
within the estimated noise between the $i$- and $z$-bands. Derived values
are in the range of 2.8--9 with many of
them close to the de Vauculeurs value of $n=4$, typical of E/S0
systems, with the exception of the object with the highest estimated redshift
(\#1446) having
$n\approx1$, that is consistent with the presence of an exponential
disk and may be an early-type spiral or perhaps an S0 galaxy. This
galaxy is reminiscent of the $z=2.5$ evolved disk galaxy discussed by
Stockton et al. (2004).
For object \#4950 the S\'ersic fit leaves strong
residuals, corresponding to the features visible in the $B$-band (see
Fig.~\ref{fig:images}). Nevertheless, the $z$-band light appears to be
dominated by a regular spheroid with $n\sim4$. A fit in the $i$-band was
not attempted for this object.

The effective radii in the $i$- and $z$-bands are also in most cases
consistent,
suggesting the lack of strong color gradients within the galaxies. 
Some of the typical effective radii are very small, 
of order $r_e\sim2$--4 ACS-UDF pixels, corresponding to 0.06--0.12$''$
for the 0.03$''$/pix
scale of the drizzled UDF images, or to a physical size $\simlt1$~kpc
(proper length).
We will come back to these issues in Sect.~\ref{sec:mor_evol}.
The axial ratios $b/a$ are also consistent between the two bands,
generally quite close to unity and consistent with the range expected 
for E/S0 systems.

The proposed redshift identifications result in a reasonably 
complete (Sect.~\ref{sec:compl}), magnitude limited sample 
of $z>1.4$ galaxies representative of 
old stars dominated, morphologically established early type galaxies.
These high-redshift spheroids appear to be already dynamically
relaxed, with the possible exception of object \#4950.

\begin{deluxetable*}{rccccccccccccc}
\tabletypesize{\tiny}
\tablecaption{Stellar populations properties}
\tablewidth{0pt}
\tablehead{
\colhead{ID} & \colhead{$z_{spec}$} & \colhead{Class} &
\colhead{$\chi^2_{r,m}$} & \colhead{P($\chi^2_{r,m}$)} & \colhead{$M_B$} &
\colhead{U-B$_{\rm rest}$} & \colhead{B-V$_{\rm rest}$} 
& \colhead{$M_*$} & \colhead{$M_*/L_B$} & \colhead{Age$_{\rm pass}$} & \colhead{$z_{\rm pass}$} 
& $S_{\rm type}$
& \mguv \\
\colhead{} & \colhead{} & \colhead{} & \colhead{} & \colhead{\%} & 
\colhead{Vega} & \colhead{Vega} & \colhead{Vega} 
& \colhead{$10^{11}M_\odot$} & \colhead{$\odot$} & \colhead{Gyr} & \colhead{} 
& \colhead{} & \colhead{}
}
\startdata
8238 & 1.39 & B & 1.5 & 13 & -20.98 & 0.23 & 0.91 & 1.0--2.4 & 2.5--6.2 & 0.5--4.5 & $>1.6$ 
& F5 & $1.14\pm0.19$ \\
4950 & 1.55 & A & 1.2 & 30 & -22.86 & 0.19 & 0.76 & 2.9--7.3 & 1.3--3.4 & 0.6--2.5 & 1.8--3.7
& F0 & $1.17\pm0.04$ \\
1025 & 1.73 & A & 0.7 & 68 & -21.76 & 0.19 & 0.76 & 1.0--1.8 & 1.3--2.2 & 0.6--1.2 & 2.1--2.6
& A3 & $1.14\pm0.06$ \\
3523 & 1.76 & B & 1.7 &  8.4 & -21.73 & 0.20 & 0.67 & 1.0--1.5 & 1.3--1.8 & 0.5--1.2 & 2.1--2.6
& F2 & $0.94\pm0.09$ \\
3650 & 1.91 & A & 1.7 & 7.8 & -22.32 & 0.20 & 0.67 & 1.3--2.0 & 1.0--1.5 & 0.6--1.6 & 2.3--3.4
& F0 & $1.25\pm0.05$ \\
3574 & 1.98 & A & 2.8 & 0.2 & -21.41 & 0.16 & 0.68 & 0.5--0.9 & 0.9--1.6 & 0.5--1.2 & 2.3--2.9 
& F0 & $1.5\pm0.2$ \\
1446 & 2.47 & B & 10 & 0 & -21.62 & 0.20 & 0.59 & 0.7--1.1 & 1.0--1.7 & 0.3--0.7 & 2.8--3.2
& F5 & $2.8\pm0.6$
\label{tab:physical}
\enddata
\tablecomments{$\chi^2_{r,m}$ is the reduced $\chi^2$ of the best fit.
Ranges for $M_*$ and $Age_{pass}$ (and thus for $M_*/L_B$ and
$z_{pass}$) are given at the 95\% confidence level. 
$U-B$ and $B-V$ colors are derived from the best fitting model.}
\end{deluxetable*}

\section{Results: stellar populations of $z>1.4$ early-type galaxies}
\label{sec:fits}

In order to derive the characteristic stellar population properties of the
seven $z>1.4$ galaxies we compared the photometric SEDs (with deep imaging
data in 10 bands from
B to K, as described in Sect.~\ref{sec:photoz}, see also
Fig.~\ref{fig:SEDs})
to stellar population synthesis
models from the Bruzual \& Charlot (2003) library, at the
(fixed) spectroscopic redshift of each source. 
We base our physical parameter estimates mainly on the photometric
SEDs rather than on the spectra because of the much wider wavelength 
range spanned by the former. In particular we would expect the rest-frame UV
photometry, including the very deep $B$-band
images, to strongly constrain the presence or lack thereof of young
stellar populations. The near-IR bands provide instead a census of the past
star-formation. 

A more extended library than the one adopted in Sect.~\ref{sec:photoz} 
was used in this case, including models with 0.2, 0.4, 1.0 and 2.5 solar 
metallicity.
Reddening is restricted to be $E(B-V)<0.2$ (we use a Calzetti et al. 2000 law). 
Some small
amount of dust reddening are sometimes found also among local early-type
galaxies (e.g., Goudfrooij \& de Jong 1996), 
and could be more often present at these high redshifts as we  are
observing more closely to the star-formation epoch. For the star-formation 
history we adopted SSP models and exponentially decreasing models with
$\tau=0.1$, 0.3, 1.0 Gyr. In addition we considered models with
step-wise star-formation history, with constant SFR lasting for 0.1, 0.3,
1 and 2 Gyr, followed by a period of
passive evolution.  Only ages less than the
age of the Universe at the fitting redshifts were allowed.
The parameters we are mainly interested in are the 
age and stellar mass of the galaxies and we determined 95\% ranges for
these two parameters following Avni (1976) and marginalizing over
$E(B-V)$, metallicity and SF history
(Table~\ref{tab:physical}). 
We also fitted main
sequence stars templates from the Kurucz library to the HST 
spectra of all galaxies (Fig.~\ref{fig:speALL}). The \mguv indexes were
measured. Indeed, in most cases they are significantly larger than
1, as expected for old stellar populations.

All the results are shown in Table~\ref{tab:physical}. 
Best fits at the spectroscopic redshifts are 
shown for all objects in Fig.~\ref{fig:SEDs}. The reduced $\chi^2_m$ are
generally low, with large associated probabilities, and the best fits to
the photometric SEDs (Fig.~\ref{fig:SEDs}) look generally very good.
An exception is object \#1446 with large $\chi^2_m$ at the 
tentative spectroscopic redshift $z=2.47$. Still, its SED looks
reasonably well reproduced at the eye inspection (Fig.~\ref{fig:SEDs}).
In part the bad  $\chi^2_m$ may be due to the presence of AGN continuum
emission starting to contribute to the near-IR bands fluxes.

The derived masses are generally large, in the range
0.5--7.3$\times10^{11}M_\odot$. Notice that these values were derived
using the total flux of these galaxies estimated from S\'ersic profile
fittings. The use of Kron magnitudes derived by SExtractor would have 
implied an underestimation of the masses by $\sim40$\% on average.
Absolute magnitudes and colors were derived for our targets by using 
the best fitting models for computing K-corrections. 
The derived $B$-band rest-frame stellar mass to light ratios are typically 
about  1--2
in solar units, as compared to the dynamically estimated total (thus including 
the contribution of dark matter) mass to light ratios
of up to 6 for local early-type galaxies (van der Marel 1991). 
Accounting for the fact that the adopted  Salpeter IMF all the way down 
to $0.1\,
M_\odot$ may be over-estimating
the stellar masses (and mass to light ratios) by at least 30\%, and that dust 
reddening in the fit can affect by up to 1 magnitude the $B$-band
rest-frame, 
this is consistent with a dimming of a few magnitudes 
from passive evolution to $z=0$ as computed from aging the best-fitting
Bruzual \& Charlot models. The inferred stellar mass to light ratios at
$z=1.4$--2 appear also consistent with the trend established for cluster
early-type galaxies up to $z\sim1.3$ on the basis of fundamental plane
studies (van Dokkum et al. 2004; Renzini 2004).
The 95\% ranges on fitted properties are not very narrow in
most cases. 
Availability of Spitzer
photometry providing access to the near-IR rest-frame could 
allow to further improve these estimates. However, the uncertainties about the 
importance and modeling of AGB stars contribution at those wavelengths
(Maraston 2005) could further complicate the issue, our sample being perhaps the
ideal one to verify the predictions of different models. A future publication 
will address these points.

\begin{figure*}[ht]
\centering 
\includegraphics[width=17cm]{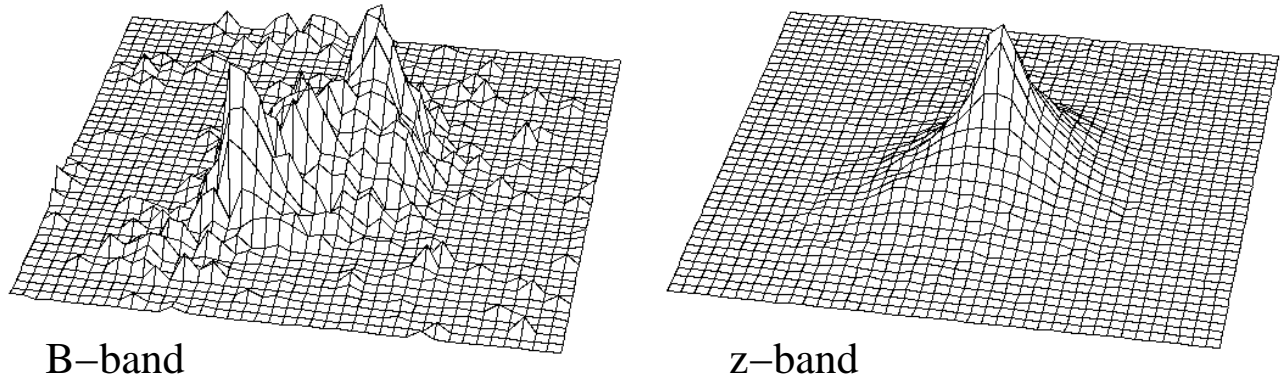}
\caption{Comparison of the surface brightness distribution for object
\#4950 in the $B$- and $z$-band. The same range of fluxes
($f_{\lambda}$) are plotted for the two objects. The peak of the
$z$-band image corresponds to a relative dip in the B-band, that it is
instead dominated by two blobs separated by $\sim1$ kpc (0.1$''$) from the 
galaxy center. The z-band surface brightness profile clearly shows the central 
cusp and extended wings typical of de Vauculeurs galaxies.
The faint spiral arms, or ring, are not visible in these images. See also
Fig.~\ref{fig:images} and the color inset in Fig.~\ref{fig:speALL} for
comparison.
}
\label{fig:both}
\end{figure*}

\subsection{Formation and ages of $z\simgt1.4$ early-type galaxies}

In order to characterize the evolutionary status of these objects, we
estimated the quantity Age$_{\rm pass}$ as the time elapsed since the
onset of the passive evolution (i.e. since the end of the last major
burst of star-formation).  This is defined as the difference between
the model age and the duration of the star-formation phase, assumed to
be equal to $\tau$ for exponential SF histories, zero for SSP models,
and the actual duration of the burst for truncated star-formation
models.  Table~\ref{tab:physical} shows that Age$_{\rm pass}$
estimates are typically in the range of 0.5--1.5 Gyr, suggesting that
the onset of passive evolution started at $z_{\rm pass}\sim2$--3.5 in most
cases.  

It is interesting to investigate what types of SF histories yield
acceptable fits to the SEDs, as this could constrain the SF timescales 
in early-type galaxies.
Limiting to the 5 objects with no $B$-band excess flux, we find that 
the SSP and $\tau=0.1$ Gyr
models provide much better fits to the SEDs than models with 
$\tau=$0.3 or 1 Gyr. Exponentially declining star-formation histories with 
$\tau\geq$0.3 Gyr 
are clearly inappropriate for these objects as they
strongly overproduce the rest-frame UV fluxes. A similar result
was found by McCarthy et al. (2004) and may seem to imply relatively 
short formation timescales (e.g. less than 0.3 Gyr). 
This appears to be consistent with the $\alpha$-enhancement 
found to be typical of elliptical galaxies  at $z=0$, 
that also suggests short formation timescales (e.g., Thomas, 
Maraston \& Bender 2002; Thomas et al. 2005).
However, the failure of models with $\tau\geq0.3$ Gyr may not be due to their
long  formation timescales, but to the exponentially
declining tail of star-formation.
Physical reasons, like the feedback from supernovae explosions or from
the onset of AGN activity (Granato et al. 2004), suggests
that in more physically motivated models the SF might be
completely stopped by such feedback.
The ``truncated" models that we considered
allow us to explore this possibility, and check whether much longer SF
timescales are consistent with our data.
In fact, these ``truncated" models produce acceptable
fits to the SEDs even for the case of 2-Gyr lasting star-formation.
In conclusion,
the available portion of the SED and the presence of  the \mguv feature
allow us to set strong constraints on  Age$_{\rm pass}$, but not on the previous 
duration of the SF activity. Rest-frame near-IR data from Spitzer could 
help constraining also this latter quantity.

The fits with $\tau=0.1$ Gyr models allow us to derive limits to the
maximum amount of ongoing SFR in these passive galaxies, as these
models have some residual SF at any time. For the 5 objects one gets
residual SFR$\simlt0.1M_\odot$yr$^{-1}$ with the $\tau=0.1$ Gyr
models. Even if continuing at this level for a full Hubble time
to $z=0$, this would increase the galaxies masses by less than 1\%. Of
course, we neglect here the possibility of completely dust extincted
SF that we would not detect anyway in the optical-IR.

Further constraint on the formation processes of early-type galaxies
can be obtained from the comparison with theoretical simulations. 
For example Hernandez \& Lee (2004) estimated the relaxation time-scales
of elliptical galaxies formed as the result of the merger of two equal
size disks, and the Press-Schecter probability of such a merger to
happen in practice. For early-type galaxies with stellar masses similar to the
ones that we derive, i.e. $M_*\sim 10^{11}M_\odot$, they suggest that if
such objects are found at $z\simgt1.6$, as we do, it is highly
unlikely that they have formed through mergers of equal-size spirals.

\subsection{Blue excess early-type galaxies}
\label{sec:blues}

We now come back to the two objects with B-band flux excess, \#4950
and \#1025.  It is perhaps surprising that Age$_{\rm pass}$ measurements for
these two objects are not distinguishable from the ones of the other redder
galaxies. However, unlike for the other five galaxies, these two
bluest early-type objects can only be fitted by the $\tau=0.3$ Gyr
models, which seem to produce the correct balance between old
and younger stars.  The residual SFRs of the
best-fitting $\tau=0.3$ Gyr models is
5--10~$M_\odot$yr$^{-1}$ for \#4950 and about half of that for \#1025.
This is revealed for \#4950 also by the detection of a weak 
[OII]$\lambda$3727 emission in the K20 survey spectrum.

These two objects have $BzK>-0.2$ because of their relatively blue $B-z$
color (see (Fig.~\ref{fig:BzK}). 
Passively evolving galaxies shortly after the quenching of SF are expected to 
show similar colors (Daddi et al. 2004b), but on the other hand the blue 
$B-z$ colors may also be due to a secondary burst of star-formation,
involving a small fraction of the total stellar mass.

This is most likely the case of the galaxy \#4950. Its morphology
changes strongly from the $B$-band to the $z$-band
(Fig.~\ref{fig:images} and \ref{fig:both}). The $B$-band light is
fairly irregular with a significant substructure of 3--4 blobs. It is
unlikely that such a clumpy structure is a bar, as these are typically
more regular. The blobs visible in the $B$-band may be instead due to
an ongoing merger or cannibalism of smaller galaxies or satellites.
The $z$-band light distribution is instead more regular and appears
typical of a spheroid.  A feature looking like spiral arms or perhaps
a ring is also detected at a distance of about 5 kpc from the center,
that is significantly less luminous than the central light both in
$B$ and z-bands.  From the spectrum we find additional evidence for
composite stellar populations. The best stellar template fit is in
fact an F0, but best fits with A5 and F2 stars are found when limiting
the fit to below or above 2800\AA\ rest frame, respectively.  The
$B$-bright clumps may have been formed or accreted after the $z$-band 
spheroid was in place.

The morphology of object \#1025 is instead regular from the $B$-band to
the $z$-band with no obvious changes. S\'ersic profile fits
were obtained for this object also in the $V$- and $B$-bands, finding
that it is consistent with an high S\'ersic index $n\sim4$ in all 
ACS bands.  There is instead a trend of decreasing effective radius
with decreasing wavelength that appears significant, possibly due to
the presence of a color gradient (Sect.~\ref{sec:mor_evol}). 
In the HST grism spectrum we find
that no change is detected when fitting templates of stars below or
above 2800\AA\ rest-frame only, and an A3 type star spectrum dominates
the light consistently in the 2000\AA$<\lambda<3600$\AA\ range.  There
is therefore no strong evidence in this object for the coexistence of
two clearly distinct stellar populations, although the fact that a
simple SSP model cannot reproduce its SED implies at least a somewhat
extended SF timescale. This object appears to be an early-type
galaxy observed very closely to some significant star-formation event,
when early A-type stars (rather than F-type stars found in all the
others passive objects) were still dominating its UV SED.  Broadhurst
\& Bouwens (2000) discussed the observational lack of known early-type
galaxies with A-type dominated stellar populations, our object being
perhaps the first known case. The relatively short lifetime of A-type
stars, of order of a few hundreds Myr at most, may be the actual
reason why these objects are rare.

The recent star-formation event may actually be the one when
most of the stars in this spheroid were formed.
Object \#1025 would then represent the closest known link between passively 
evolving F-star dominated early-type galaxies and the as yet not
established nature of the main formation event. 
The relaxed morphology at this early stage would
imply a fairly short dynamical relaxation timescale.
If the star-formation event is a secondary one, involving only a
fraction of the mass, this objects might instead be
similar to the E$+$A galaxies found at lower redshift (Dressler et al.
1983). 
This single object in the $1.4<z<2$ redshift range
corresponds, however, to a volume density that is 
2 to 3 orders of magnitude higher than the local density of E$+$A
galaxies above similar luminosities (Quintero et al.  2004). The 
fraction of E$+$A galaxies in the present sample (one out of seven)
would also be higher then the just 
1\% that is found locally (Quintero et al.  2004).

\section{Results: evidence for a declining number density of early-type 
galaxies at $z\simgt1.4$}
\label{sec:evol}

An important question to address is how the abundance of $z\simgt1.4$ 
passive early-type galaxies compares to the local value, and to the 
populations of star-forming galaxies at the same redshifts.

As the majority of our sample galaxies, and all of the class "A" ones, 
are located at $z<2$, we first consider the $1.39<z<2.00$ redshift range
for this comparison.
Over the 12.2 arcmin$^2$ selection area this  redshift
range includes  a volume of $\sim20000$ Mpc$^3$.
Summing up the galaxy stellar masses estimated in Sect.~\ref{sec:fits}
we obtain a total stellar mass in the range of 7--16$\times10^{11} M_\odot$,
and a stellar mass
density in the range of 3.4--$8.1\times10^{7} M_\odot$Mpc$^{-3}$ 
for an average redshift $<z>=1.7$.
The 6 early-type galaxies with $1.39<z<2.00$ correspond to a 
number density of $3.4\times10^{-4}$ Mpc$^{-3}$. 

\subsection{Evolution from $z\sim 2$ to $z=0$}

The masses of these 6 galaxies are all close to or larger than
$10^{11}M_\odot$, and we can
assume our sample to be  reasonably complete above such a mass threshold
for passive early-type galaxies down to $K_{AB}=23$.
In the local universe, the number density and the 
stellar mass density of passively evolving galaxies with
$M_*>10^{11}M_\odot$ are about $9\times10^{-4}$ Mpc$^{-3}$ 
and $1.3\times10^{8} M_\odot$Mpc$^{-3}$, respectively (Baldry et
al.  2004\footnote{We account for the fact that our pure Salpeter 
IMF gives stellar masses at least 30\% larger than for the IMF used in
Baldry et al. 2004}). Thus our
sample of early type galaxies in $1.39<z<2.00$  accounts, at face
value, for 
$\sim 35\%$ of the number density of $M_*>10^{11}M_\odot$ 
passive galaxies in the local universe, 
and for a fraction of $\sim 25$--60\% of the relative stellar mass density.
Notice that at $z=0$ about 75\% of all $M_*>10^{11}M_\odot$ galaxies are 
classified as passive early-type galaxies (Baldry et al. 2004). 
We therefore infer at $z=1.7$ a decrease in 
the number density of massive early-type galaxies by at most of a factor of
3, if limiting to $z<2$. 
The inferred fractions would be further divided by a factor of 2 if using
all the volume in $1.39<z<2.47$.

A major limit of these calculations is represented by cosmic variance,
because of the small volume probed by UDF.  All our passive galaxies
are EROs with $R-K>5.3$ (Table~\ref{tab:data}), which are strongly
clustered populations (Daddi et al. 2000; McCarthy et
al. 2001). Clustering can produce strong fluctuations in the number
density of objects over small volumes, which most likely result in an
underestimate of the true densities as averaged over large volumes
(Daddi et al. 2000). If one assume that these $z>1.4$ galaxies have
the correlation length of EROs, i.e. $r_0\sim10$ \h1 Mpc (Daddi et
al. 2001; 2003), one finds that even at the $1\sigma$ level the true
number density (and thus the stellar mass density) of passive galaxies
could be within half and twice that estimated from our sample, or between
20\% and  80\% of the local value.

It is clear that similar searches on much larger and independent areas 
would be important to measure the number density evolution with some
reasonable accuracy. The UDF dataset is likely to remain unique in its
depth and quality for many years to come, but our result confirms that
reliable estimates on the abundance of $z>1.4$ passive early-type galaxies
might be obtained by
using the  $BzK$ photometric criteria only
(Fig.~\ref{fig:BzK}; Daddi et al. 2004b), even without complete spectroscopic
follow-up. In fact, in the UDF the sky density of the brightest
($K_{Vega}<20$) candidate $z>1.4$ 
early-type galaxies satisfying conditions $z-K>2.5$ and $BzK<-0.2$ 
is not much smaller
than that found in a $\simgt30$ times wider  field survey (Kong
et al. in preparation). So, the densities derived over the UDF field 
should not dramatically underestimate the true values.

\subsection{Passive objects at even higher redshifts?}

The highest redshift passive object that we recover is at $z=2.47$, 
although with "B" class redshift, while all
the others lie at $z<2$. Therefore, it is important to 
ask is whether the paucity of $z>2$ passive galaxies is due 
to the limiting magnitude of the sample (i.e. they exist but are 
fainter) or to passive
galaxies getting rarer and rarer. Indeed, some of the galaxies in our sample 
are consistent with having started pure passive evolution
only at $z\simlt2$. 

We have used a $V/V_{\rm max}$ test to try to shed light on the issue.
For each of our objects we compute the maximum redshifts at which it
would still be part of the sample (i.e. with $K_{AB}<23$, Kron
magnitudes, and $z-K>2.5$). We use the best fitting Bruzual \& Charlot
models to compute K-corrections, and $z=1.39$ as the lower redshift
boundary. The result is that the six objects would be still in the
sample at least up to $z=3.3$ (when \#8238 would drop out) and up to
$z=4.6$ (when finally also \#3650 would drop out). With
$<V/V_{\rm max}>=0.14\pm0.04$ and a maximum $V/V_{\rm max}=0.35$ for \#1446, an
uniform $V/V_{\rm max}$ distribution is rejected at the 99.7\% level,
based on a Kolmogorov-Smirnov test.

There is therefore strong evidence for evolution, especially beyond $z=2$.
This means that if galaxies with masses 
and SEDs (i.e. ages) similar to the ones that we see at $<z>=1.7$ were 
present at higher redshifts we would have detected them out to $z\sim4$
(and with  $z-K>2.5$). These results
underscore a rapid disappearance of truly passive systems at redshifts
$z\simgt2$. This does not mean that massive galaxies where not present
at earlier epochs, but that if existing they were most likely still actively
star-forming.
Notice that at $z=4$ the Universe is already $\sim1.6$ Gyr old and in principle
there would have been  time to produce passively evolving
galaxies with  ages similar to the ones we see at $1.4<z<2.5$.

Of course, clustering can bias also this $<V/V_{max}>$ test. For
example, by chance the UDF may lack a cluster-like structure with many
similar galaxies at $z>2$ or 3. On the other hand, 
a similar $<V/V_{max}>$ test in the sky contiguous K20
survey region suggests that
the $K$-band luminous $BzK$-selected starbursts galaxies are evolving
positively in number over the same redshift range (Daddi et
al. 2004b). These $BzK$ starbursts are massive (stellar masses
$\sim10^{11}M_\odot$) and plausible candidates for being precursors
to early-type galaxies. Therefore it appears that while the number
density of passive galaxies is dropping with increasing $z$, that of
their immediate likely precursors is somewhat increasing, as expected.
It is worth emphasizing that these two $<V/V_{max}>$ tests were
performed basically on the same sky area, and therefore they should be
affected by clustering in a similar way.

\subsection{The fraction of $z>1.4$ stellar mass in passive objects}

It is interesting to estimate what fraction of the mass, or of the
most massive galaxies, is in passive galaxies at the probed redshifts
$1.4<z<2.5$.  Galaxies with $BzK>-0.2$ are in most cases star-forming
objects at $1.4<z<2.5$ (Daddi et al. 2004b), thus spanning the same
redshift range of the detected passive galaxies.  We can thus simply
compare the sample of $BzK>-0.2$ galaxies selected to $K_{AB}=23$
to the current sample of passive objects to the same limiting
magnitude. We used the (SED-fit based) $K$-band magnitude versus stellar
mass relation calibrated in Daddi et al. (2004b). For the passive
galaxies in our sample this relation provides fairly good estimates of
the stellar masses, fully consistent with those listed in
Table~\ref{tab:physical}.  As star-forming galaxies in $1.4<z<2.5$ we
consider all $BzK>-0.2$ sources, i.e. 43 objects when
excluding the 2 passive "blue" objects
and any source with X-ray detection as likely an AGN.
Of these 6 have estimated
stellar mass $M_*\simgt10^{11}M_\odot$, 
producing a total stellar mass
of order of $5\times10^{11} M_\odot$, or up to $10\times10^{11}
M_\odot$ for those above an estimated stellar mass limit of
$M_*>0.5\times10^{11}M_\odot$\footnote{The stellar masses estimated
for the SF
galaxy candidates at $1.4<z<2.5$ are not based on magnitudes
obtained from surface brightness profile fitting, 
as for the passive galaxies, and may thus
still be somewhat underestimated}. These figures are similar to the
ones derived above for passive objects in the same redshift and mass
ranges.  This supports earlier hints  that at
$1.4<z<2.0$ roughly $\approx50$\% of the stellar mass 
is in passively evolving galaxies and roughly $\approx50$\% is in
vigorous star-forming galaxies, for objects with stellar masses
$M_*\simgt10^{11}M_\odot$ (Cimatti et al. 2004; McCarthy et al. 2004).  
By summing together the similar stellar mass densities contributions from
passive and star-forming galaxies at $1.4<z<2$ having
$M_*\simgt10^{11}M_\odot$, one gets fairly close to the local
value from the SDSS (Baldry et al. 2004). However, the uncertainties
due to small number statistics and clustering, and in the estimates of
stellar masses are both at least a factor of $\sim2$. 
The evolution of most massive galaxies 
is not easily detectable up to $z=2$, and
therefore it is arguably not very strong.

\begin{figure}[ht]
\centering 
\includegraphics[width=8.8cm]{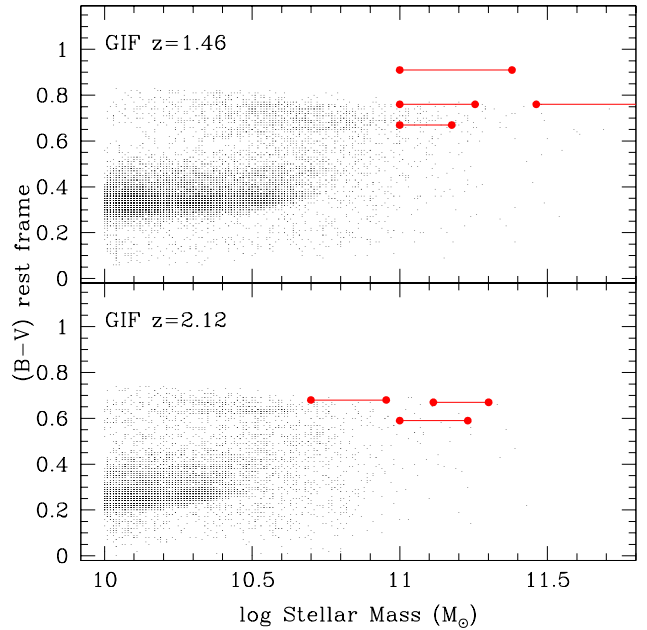}
\caption{We show the predicted stellar masses versus rest frame $(B-V)$
colors of galaxies from the $\Lambda$CDM simulations 
(Kauffmann et al. 1999). The connected large symbols are for our
objects, where minimum and maximum stellar masses are shown. The
early-type galaxies at $z<1.8$ are plotted over the $z=1.46$
simulation, while the ones at $z>1.8$ over the $z=2.12$ simulation.
The volume of each simulation is 200 times larger than the UDF
between $1.4<z<2.5$.
}
\label{fig:GIF}
\end{figure}

\subsection{Comparison to $\Lambda$CDM theoretical predictions}

 McCarthy et al. (2004) and Cimatti et al. (2004) already pointed
out
 that the existence of $z>1.4$ passive early-type galaxies in
such
 relatively high number is at odd with current predictions of
semianalytic models of galaxy formation and evolution, that predict
the passive evolution phase to be established much later, if
ever. This is similar to the reported failure to accounts for EROs
number counts even at $z\sim1$ (e.g., Daddi et al. 2000; Firth et
al. 2002).  These conclusions are supported also by the present
findings.
 
 For example, among the {\it early} semianalytic models,
the
 simulations\footnote{http://www.mpa-garching.mpg.de/GIF/} 
 by
Kauffmann et al. (1999) predict a number density at $z=1.46$ of
galaxies with $(B-V)_{rest}>0.6$ (Table~\ref{tab:physical})
 and
stellar masses $M_*>10^{11}M_\odot$ that is about a factor of 10 lower
than recovered here in the UDF, and by $z\sim 2$ such objects should
have virtually disappeared. Nevertheless, it is interesting that these
models
 do predict that at least some galaxies exist with the colors
and masses of the
 ones we observe (Fig.~\ref{fig:GIF}), and that a
hint for a bimodal color
 distribution is already in place in the
models 
 at these redshifts. On the other hand, there appear to be too
few
 massive galaxies in these simulations, either passive or star
forming
 (Daddi et al. 2004a,b; Fontana et al. 2004). Among the more
recent generation of models, those by Somerville et al. (2004) are
quite successful in predicting the number density of massive galaxies
at high redshift, but fail to produce the bimodal color distribution
even at low redshift.  Similarly, the hydrodynamical simulations of
Nagamine et al. (2005a;b) appear to match the statistic of $z=2$
massive galaxies but are less successful in reproducing the abundance
of $z\sim2$ old and passive galaxies.  There is now general agreement that one of the problems with the models is their tendency to sustain star formation in massive DM halos all the way to low redshifts, and  intense
theoretical efforts are currently under way in the attempt to cope
with these discrepancies. A strong AGN feedback appears to be  a
viable way for switching off star formation in massive galaxies at
$z\simgt 2$ (e.g., Granato et al. 2004).

\section{Results: morphology evolution of early type galaxies
to $z>1.4$}
\label{sec:mor_evol}

The morphological properties of our sample of $z>1.4$ passive early-type
galaxies were compared to those of the corresponding local  populations.
\begin{figure}[ht]
\centering 
\includegraphics[width=8.8cm]{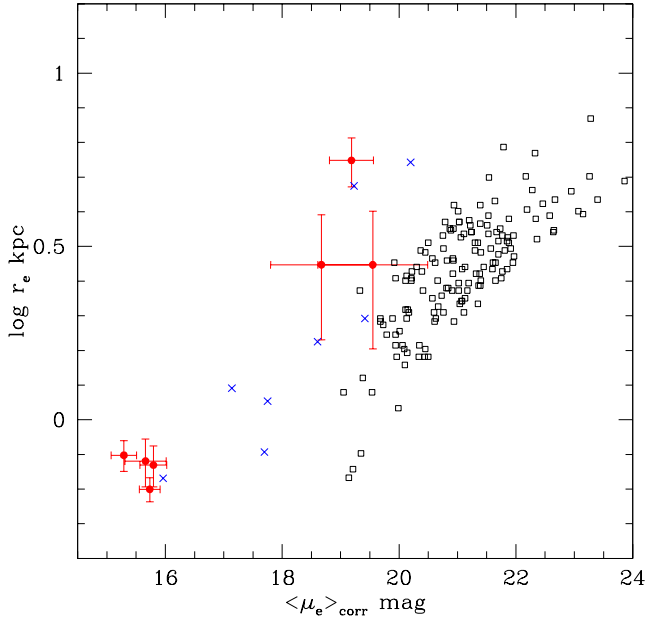}
\caption{The B-band rest-frame Kormendy (1977) relation 
for the $1.4<z<2.5$ early-type
galaxies (filled circles with error bars), 
compared to Coma early type galaxies taken from Joergensen et
al. (1995) converted to the B-band (empty squares), and to 
$0.7<z<1.2$ early-type galaxies from the K20 survey (crosses,
from di Serego Alighieri et al. 2005, in preparation).
Observed values for our objects are derived from the $z$-band
measurements.
The average effective brightness $<\mu_e>$
were corrected for cosmological surface brightness dimming 
and bandpass
shift (observed $z$-band to rest $B$-band), but not for the 
(relatively uncertain) passive evolution dimming to $z=0$.
}
\label{fig:Korme}
\end{figure}

Fig.~\ref{fig:Korme} shows the B-band rest-frame 
Kormendy (1977) relation for our
objects compared to a local sample of early-type galaxies in the Coma
cluster (Joergensen et al. 1995) and to a sample of massive 
$z\sim1$ early-type galaxies from the K20 survey (di Serego
Alighieri et al. 2005, in preparation). The average effective surface
brightness $<\mu_e>$ (see e.g., Ziegler et al. 1999 for its definition)
has been corrected to $z=0$ accounting for the $(1+z)^4$ cosmological
surface brightness dimming. A color term was applied to transform our
$z$-band measurements to rest-frame $B$-band, and similarly the
Joergensen et al. (1995) $r$-band measurements were corrected using
$B-r=1.15$ (Ziegler et al. 1999). When not accounting for the expected
dimming of the $B$-band rest-frame due to the aging
of the stellar populations, the $z>1.4$ objects are largely offsetted
from the local Kormendy relation (Fig.~\ref{fig:Korme}).  The expected
dimming to $z=0$ can be computed on the basis of the inferred
best-fitting Bruzual \& Charlot (2003) models, although with large
uncertainties given to the range of acceptable best fitting parameters
(especially ages and SF histories).  Typically a dimming by a few
magnitudes is derived, and these large passive evolution corrections,
albeit uncertain, would bring the 7 passive $z>1.4$ galaxies in
reasonable agreement with the local Kormendy relation.

We notice, however, that 4 out of the 7 objects have very small
effective radii, $r_e<1$ kpc, and passive evolution would bring them
in a region of the $r_e-\mu_e$ diagram where very few local galaxies
are found.  Moreover, local galaxies with $r_e<1$ kpc tend to be much
less massive than our $z>1.4$ objects.  
Similarly, the stellar surface mass density
$\mu_*$ of the 4 galaxies, defined as $\mu_* = M_*/(2r_e^2)$
is more than a factor $\sim 10$ higher than that of local
galaxies with similar stellar masses (Kauffmann et
al. 2004). All in all, the 
$z>1.4$ $r_e<1$ kpc objects cannot be the progenitors of the $z=0$, 
low-mass galaxies with similar effective radii.

Similarly, about half
of $z\sim1$ massive early-type
galaxies from the K20 survey, shown in Fig.
\ref{fig:Korme}, have sizes of about 1 kpc that are 
smaller then their local massive counterparts.
Small sizes of K20 early-type galaxies 
were also inferred by Cassata et al. (2005) 
for the higher redshift  
sample of 
4 early-type galaxies at $z>1.5$, i.e., $r_e=1$--3 kpc, 
measured on ACS
$z$-band data as in this work. 
Waddington et al. (2002) showed that
their two $z\sim1.5$ radio-selected early-type galaxies have too small sizes (3 kpc), 
as compared to lower redshift samples. Stanford et al. (2004)
present a rest-frame $B$-band morphological analysis of early-type
galaxies to $z=1.4$ in the Hubble Deep Field North based on HST+NICMOS,
showing indeed a trend of significantly smaller sizes of luminous
galaxies with respect to local massive counterparts (see
their Fig.~16). An
analogous result might be noticed in Fig.~4 of Gebhardt et al.
(2003). 

Therefore, it appears well established
observationally that some fraction of high redshift passive early-type 
galaxies appear significantly smaller then their likely
$z=0$ descendants. While this is similar to the well known trend of size 
reduction with redshift for star-forming galaxies (e.g., Ferguson et al. 
2004), this effect had not been
highlighted and/or discussed previously for early-type galaxies, if not for the
2 radio-galaxies of Waddington et al. (2002).
It is somehow surprising because one would expect that these massive 
galaxies, that are already undergoing passive evolution,
should evolve smoothly into 
local massive ellipticals with no major change in their properties if not
for aging of the stellar populations.
The availability of the ultradeep UDF imaging might allow us
to better constrain and understand this issue.

We tried to investigate the possible reasons for this
effect on our sample of $z>1.4$ early-type galaxies.
The small effective radii we derive with the S\'ersic 
fitting might be
due to the presence of an unresolved nuclear component, e.g., from an
AGN.  Actually, two of the four objects with small $r_e$ are detected
in X-rays (see Section \ref{sec:Xrays}).  For the two compact objects
with the highest S/N ratio in the $z$-band
imaging (\#1025 and \#3650) we attempted
fitting their surface brightness profiles with two component models, 
S\'ersic plus a point-like source.  The
resulting best fit is still dominated by the S\'ersic component, but
its $r_e$ is not significantly increased with respect to the one
component fits, just a lower S\'ersic index $n$ is derived.  However,
when forcing $r_e$ to be large, a good fit with $r_e\simeq$few kpc can
be obtained, consistently with the local objects, and still with large
$n$ values. In this case the nuclear source is about 2 magnitudes
fainter than the spheroid, in agreement with the fact that the optical
SEDs and spectra appear dominated by stellar light
(Sect.~\ref{sect:reds} and
\ref{sec:fits}). 
If this is the reason of the small radii, it would imply a widespread
nuclear activity, or relics of nuclear starbursts,
in $z\sim1.4$--2 early-type galaxies. Close inspection of the inner
regions of local early-type 
galaxies does in fact  reveal the presence of faint nuclear point sources 
in nearly 50\% of the cases (Ravindranath et al. 2001). Those 
nuclear sources might have been significantly more luminous/frequent at high 
redshifts.

\begin{figure}[ht]
\centering 
\includegraphics[width=8.8cm]{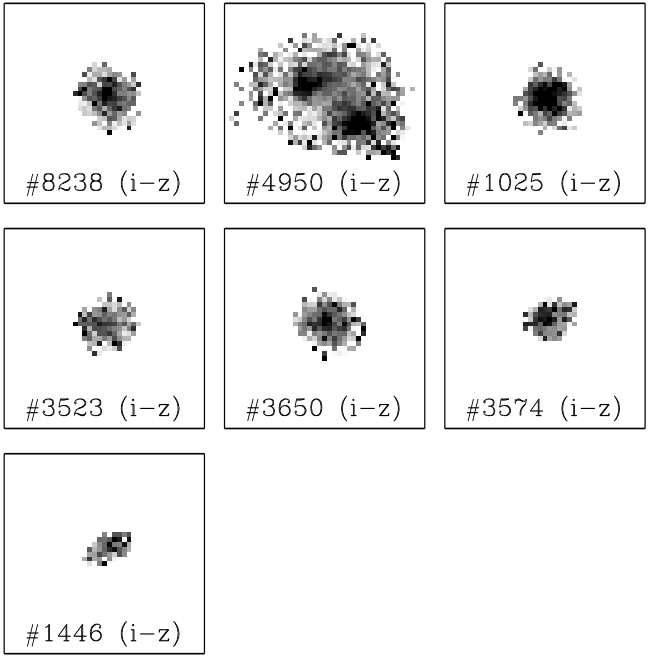}
\caption{The $(i-z)$ color maps: darker regions correspond to bluer $i-z$ 
colors. Only pixels with S/N$>3$ in both $i$ and
$z$ bands were considered. No attempt was made to account for the
different PSF between the two bands, so that the bluer (i.e. darker in
the images) cores of most objects are likely the result of PSF
differences. Obvious exceptions are object \#4950 where 2 blue blobs are
clearly visible, and \#1025 that shows a strong color gradient and blue
core.
}
\label{fig:CM}
\end{figure}

If instead the small $r_e$ values are not (or not always)
due to the presence of
nuclear emission, two possibilities may help to solve the puzzle, and
reach consistency with the local galaxy populations: either a strong
morphological K-correction or the physical sizes will increase with
time. Larger sizes at $z=0$ would move the galaxies roughly along the
Kormendy relation by decreasing the average effective surface
brightness, and would bring our objects positions in
Fig.~\ref{fig:Korme} in better agreement with local ones.

A possible mechanism for size evolution would be substantial satellite
engulfment by the massive red galaxies as they evolve to lower
redshifts, which would imply an additional growth of their stellar
mass. Object \#4950 might be an example of this process in
act. Confirmation of this scenario would require the detection of
large concentrations of satellite galaxies in the vicinity of the passive
$z>1.4$ galaxies. This option might be tested in the future by
using the {\it ultradeep} UDF data.

\begin{figure}[ht]
\centering 
\includegraphics[width=8.8cm]{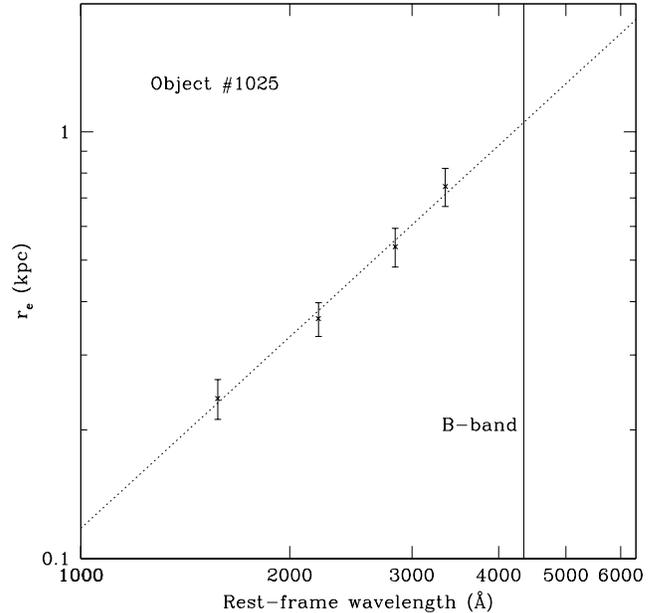}
\caption{The dependence of the effective radius on wavelength for object
\#1025 at $z=1.73$. The observed trend would imply $r_e$ about 50\% larger at
rest-frame B-band (4350\AA)
with respect to observed z-band (or rest-frame 3350\AA), but still
$\sim1$ kpc only.
}
\label{fig:re_w}
\end{figure}

In relation to the morphological K-correction option, we note that
blue cores are not uncommon already among $0.5<z<1$ ellipticals (Menanteau et
al. 2001, 2004), and that the observed $z$-band corresponds to
2600--3800\AA\ in the rest frame for $1.4<z<2.5$ (so somewhat bluer
than the 4300\AA\ of the $B$-band).  The most direct way to check this
option would be to obtain S\'ersic profile fits in the NICMOS F110W
and F160W bands, bracketing the rest-frame $B$-band for
$1.4<z<2.5$. However, this is technically difficult, mainly due to the
large 0.2$''$ pixel size of NICMOS (a factor of 4 larger than the ACS
pixels) corresponding to $\sim 2$ kpc at $z\sim 2$. Attempting such
fits is beyond the scope of the present work. What we checked instead
was the trend of $r_e$ with wavelength, within the range covered by the
ACS data. We constructed $i-z$ color maps of all our $z>1.4$ galaxies
(Fig.~\ref{fig:CM}), but among the 4 objects with $r_e<1$ kpc only
\#1025 shows a quite prominent blue core. This objects also exhibits a
strong trend of $r_e$ with wavelength, as shown in
Fig.~\ref{fig:re_w}, while a S\'ersic profile consistent with
$n\sim4$ is found in all the ACS bands. An extrapolation would give $r_e$
larger by 50\% in the rest-frame $B$-band with respect to that
measured in the observed $z$-band. Still, this increase would not
suffice to bring this object at $z=0$ on the populated part of the
Kormendy relation, as that would need an increase by a factor $\sim
3$. Yet, we cannot presently rule-out the possibility that $r_e$
may change more steeply with wavelength, especially beyond the
4000\AA\ break.  We finally note that small $r_e$ values in the
rest-frame $B$ band are common among early-type galaxies to $z=1.4$
(Stanford et al. 2004), which outlines the existence of this problem
also when using B-band rest-frame measurements. 
A final assessment of why some high-$z$ ellipticals
have such small effective radii and how they relate to local E/S0
galaxies is left to future investigations.

\section{Results: the X-ray properties of $z>1.4$ early-type galaxies}
\label{sec:Xrays}

\begin{deluxetable*}{cccccccccc}
\tabletypesize{\scriptsize}
\tablecaption{X-ray emission properties}
\tablewidth{0pt}
\tablehead{
\colhead{ID} &  \colhead{ID(G02)} & \colhead{ID(A03)} & \colhead{$z$} & 
\colhead{$F_{soft}$} & \colhead{$F_{hard}$} & \colhead{$N_H$} & 
\colhead{$L_{2-10 keV}$} & \colhead{X/O} & \colhead{X/K}\\
\colhead{ } & \colhead{ } & \colhead{ } & \colhead{ } & 
\colhead{cgs} & \colhead{cgs} & \colhead{cm$^{-2}$} & 
\colhead{cgs} & \colhead{} & \colhead{ }
}
\startdata
\label{tab:X}
1025 & 256 & 255 & 1.73 & 
$0.9\times10^{-16}$ &  $4.1\times10^{-15}$ & $\sim5\times10^{23}$ 
& $8\times10^{43}$ & 12 & 2.5 \\
1446 & 605 & 243 & 2.47 & 
$<0.4\times10^{-16}$ &  $4.6\times10^{-16}$ & (HR$>0.42$) &
$2\times10^{43}$ & 5 & 1 \\ 
\enddata
\end{deluxetable*}

Two of the 7 $z>1.4$ early-type galaxies discussed here are detected
in the 1 Msec Chandra X-ray observation (Giacconi et al. 2002), and
their properties are reported in Table~\ref{tab:X}.  The two sources
are both very hard ($HR>0.4$) and both of them were discussed also by
Padovani et al. (2004) as possible QSO2 candidates\footnote{The
luminosities derived by Padovani et al.  (2004) are larger than those
inferred here given that these authors put these objects at
slightly higher redshifts.}.

\begin{figure}[ht]
\centering 
\includegraphics[width=9.8cm,angle=0]{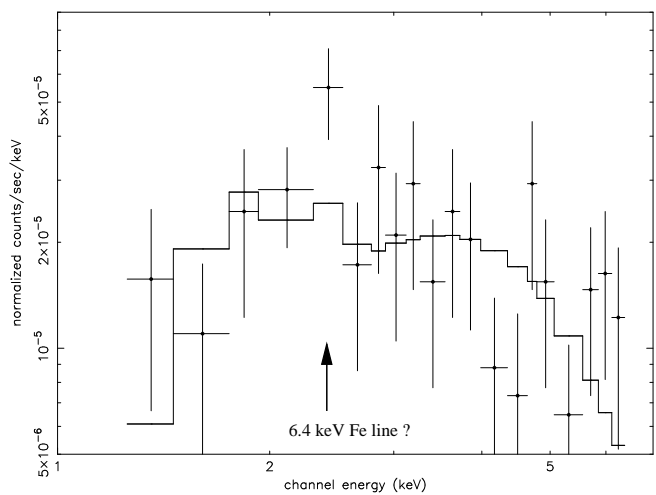}
\caption{The Xray spectrum of \#1025 from Chandra, fitted with a 
$\Gamma=2.0$ power law and absorbing column density of
$N_H=5\times10^{23}$~cm$^{-2}$. The excess flux at 2.4 keV is consistent
with the 6.4 keV Fe line, redshifted at $z=1.73$.
The presence of such a strong Fe emission is reasonable (Brusa et al. 2005b),
given also the very hard X-ray spectrum and high absorption column density
of the source.
}
\label{fig:1025_X}
\end{figure}

Object \#1446 is close to the detection limit with $\sim20$ net
counts, almost all from the hard band (HR$>0.42$), and we estimate its
luminosity to be L$_{2-10 \rm keV}=2.2\times10^{43}$ erg s$^{-1}$.
Object \#1025 is detected with $\sim90$ net counts, thus allowing a
rough spectral analysis.  Fig.~\ref{fig:1025_X} shows the X-ray
spectrum of the object along with an absorbed power-law at the source
redshift.  The best fit rest frame column density id $N_{\rm
H}$=5$\times10^{23}$ cm$^{-2}$; and the corresponding unabsorbed
X--ray luminosity is L$_{2-10 \rm keV}$=8$\times10^{43}$ erg s$^{-1}$.
Given the high column density and the X--ray luminosity close to
10$^{44}$ erg s$^{-1}$ this object can be classified as an X--ray
QSO2. It is intriguing that the optical spectrum appears to be
dominated by stellar light, that may imply that the obscuring torus is
blocking also most of the emission in the optical and near-IR domains.
An excess emission with respect to the power-law continuum is present
at $\sim 2.3-2.4$ keV.  This would be consistent with the presence of
a redshifted 6.4 keV Fe K$\alpha$ emission line. Indeed, when a
Gaussian line is added to the model spectrum, the fit
improves and suggests a redshift $z=1.71$, in excellent agreement with
the redshift measured in the optical/near-IR, and a rest-frame line
equivalent width of $\sim350$ eV, similar to what found in obscured
X-ray sources (e.g., Brusa, Gilli \& Comastri 2005b; Maccacaro et al. 2005).

Both sources have large X--ray to optical 
flux ratios (X/O). Coupled with the analysis of Mignoli et al. 
(2004), that suggest that most of the hosts of hard X-ray AGN with 
large X/O and R-K$>5$ are spheroids at $z>1$, this imply a close 
connection between hard X-ray sources and high-redshift early-type galaxies.

Although based on two  objects only, the fraction of AGNs in $z\sim2$
these early-type galaxies is at face value $\sim 30\%$. 
This is larger than that observed for the early--type sample 
of $z\sim1$ EROs discussed by Brusa et al. (2002; no X--ray emission
was revealed in a sample of 8 sources), and much larger than 
the $\sim1$\% value observed locally. 
This is similar to the fraction
of AGN among EROs at the probed K magnitudes (Alexander et al. 2002;
Brusa et al. 2005a) underlinyng again the possible
close connection between the formation of AGN and early-type galaxies.

\section{Summary and conclusions}
\label{sec:summary}

Based on the $BzK$ two-color selection criterion meant to identify $z>1.4$
passively evolving galaxies, a sample of seven objects has been identified
in the UDF for which ACS grism spectra confirm both the high redshift
($1.4\simlt z\simlt 2.5$) and the current absence (or very low) ongoing star
formation. The  analysis conducted over the {\it ultradeep} ACS imaging
further shows that the objects are morphologically early-type galaxies.

The redshifts are derived from the identification of the spectral feature 
at the rest-frame $2640<\lambda <2850$ \AA\ due mainly to MgI and MgII
absorptions in late A- and  F-type stars, and which shows up prominently
only in synthetic stellar populations with no (or very low) star formation
in the last $\sim 0.5$ Gyr.

The SEDs and ACS grism spectra of 5 of the objects are best reproduced by a
star formation history initiated $\simgt 1.5$ Gyr before, and
completely discontinued over the last $\sim 0.5-1.5$ Gyr, implying that
the passive evolution started at $z\simeq 2-3.5$. In the  other
cases a similar age constraint is inferred, but the spectra are best
reproduced by an exponentially declining SFR, with $\tau\sim 0.3$ Gyr.

The overall spectral energy distributions of the objects indicate
stellar masses in excess of $\sim 10^{11}M_\odot$ and we infer that
the comoving volume density of such massive, early-type galaxies at
$<z>\simeq 1.7$ is about $1/3$ of the local value. The uncertainty in
the derived fraction is dominated by cosmic variance, given the small
field covered by the UDF. Allowing for the expected clustering of
these galaxies we infer that at the $1\sigma$ level such fraction
could be as low as $\sim 20\%$ or as high as $\sim 80\%$. Clearly, the
exploration of wider fields is necessary to accurately pinpoint the
actual decrease with redshifts of the number density of massive
galaxies which are passively evolving.  However, a $V/V_{\rm max}$ test
indicates that beyond $z\sim 2$ this number density is likely to drop
very rapidly.

One intriguing aspect of the present findings is the quite small
effective radii derived from the ACS $z$-band images, which in 4 out
of 7 cases are less than 1 kpc. Given their high mass, the passive
evolution of such very compact objects will bring them in a region of
the Kormendy relation (or, equivalently, of the fundamental plane)
which is depopulated at $z=0$. We discuss various possibilities to
explain this apparent paradox, including a morphological K-correction
(e.g., due to the possible presence of blue cores in these galaxies),
or of an AGN point-like source biasing the $r_e$ measurements, or
eventually some evolutionary effects such as if the observed galaxies
were still subject to further growth at lower redshifts. While we mention
hints favoring one or another option, the existing data do not allow to
reach any firm conclusion on this specific issue.

Finally, we point out that while in the local universe  most of the
most massive galaxies are passively evolving giant ellipticals, the present 
data -- combined with previous evidence in a partly overlapping field
(Daddi et al. 2004a;b) --
indicates that by $z\sim 2$ passively-evolving and seemingly vigorous
starbursts galaxies
occur in comparable numbers among most massive galaxies. 
This remains a major challenge for the current 
theoretical simulations of galaxy formation to reproduce.


\acknowledgments
We thank Swara Ravindranath,
Ignacio Trujillo and Paolo Cassata for
providing comments about the morphological analysis,
Andrea  Comastri for help with Fig.~\ref{fig:1025_X}, and Eros Vanzella
for discussions. The anonymous referee is thanked for a careful reading
of the manuscript and useful comments. Support for part of this work
was provided by NASA through the Spitzer Fellowship Program, under
award 1268429. We acknowledge support from the grant GO-09793 
from the Space Telescope Science Institute, which is operated by AURA under
NASA contract NAS5-26555. This project has made use of the aXe
extraction software, produced by ST-ECF, Garching, Germany.

\citeindexfalse

\end{document}